\documentclass[epsfig]{elsart}
\usepackage{amsfonts,amssymb,graphicx}
\begin{document}

\newcommand{\re}{\mathop{\mathrm{Re}}}
\newcommand{\im}{\mathop{\mathrm{Im}}}
\newcommand{\D}{\mathop{\mathrm{d}}}
\newcommand{\I}{\mathop{\mathrm{i}}}
\newcommand{\E}{\mathop{\mathrm{e}}}

\noindent {\Large DESY 05-164}

\noindent {\Large September 2005}

\begin{frontmatter}

\journal{Optics Communications}
\date{}

\title{
Properties of the odd harmonics of the radiation from
SASE FEL with a planar undulator
}

\author{E.L.~Saldin},
\author{E.A.~Schneidmiller},
and
\author{M.V.~Yurkov}

\address{Deutsches Elektronen-Synchrotron (DESY),
Hamburg, Germany}

\begin{abstract}

Recent theoretical and experimental studies have shown that
Self-Amplified Spontaneous Emission Free Electron Laser (SASE FEL) with
a planar undulator holds a potential for generation of relatively
strong coherent radiation at the third harmonic of the fundamental
frequency. Here we present detailed study of the nonlinear harmonic
generation in the SASE FEL obtained with time-dependent FEL simulation
code. Using similarity techniques we present universal dependencies for
temporal, spectral, and statistical properties of the odd harmonics of
the radiation from SASE FEL. In particular, we derived universal
formulae for radiation power of the odd harmonics at saturation. It was
also found that coherence time at saturation falls inversely
proportional to harmonic number, and relative spectrum bandwidth
remains constant with harmonic number.

\end{abstract}

\end{frontmatter}

\clearpage
\setcounter{page}{1}

\section{Introduction}

Radiation of the electron beam in the planar undulator contains rich
harmonic spectrum. This referes to both, incoherent and coherent
radiation as well. During last years a significant efforts of
researchers have been devoted for studying the process of the higher
harmonic generation in the high-gain free electron lasers
\cite{hg-1}-\cite{3rdharm-ttf}. Such an interest has been mainly driven
by practical needs for prediction of the properties of X-ray free
electron lasers. A fraction of a higher harmonic content is very
important for users planning experiments at X-ray FEL facility. On the
one hand, higher harmonics constitute rather harmful background for
some classes of experiments. On the other hand, higher harmonic
radiation can significantly extend operating band of the user facility.
In both cases it is highly desirable to know properties of the hiher
harmonic radiation. Analytical techniques have been used to predict
properties of the higher harmonics for FEL amplifier operating in the
linear mode of operation \cite{kim-1,kim-2}. However, the most fraction
of the radiation power is produced in the nonlinear regime, and a set
of assumptions needs to be accepted in order to estimate saturation
power of higher harmonics on the base of extrapolation of analytical
results. As for statistical properties, they could not be extrapolated
from linear theory at all. A lot of studies has been performed with
numerical simulation codes. These studies developed in two directions.
The first direction is investigations of higher harmonic phenomena by
means of steady-state codes \cite{hg-3,hg-4,hg-5,hg-6}. Despite the
results of these studies are applicable to externally seeded FEL
amplifiers, it is relevant to appreciate that they gave the first
predictions for high radiation power in higher harmonics of SASE FEL.
Another direction was an extraction of time structure for the beam
bunching from time-dependent simulation code with subsequent use of
analytical formulae of the linear theory \cite{kim-1}. Giving an
estimate for the power, such an approach does not allow to describe
statistical properties of the output radiation.

In this paper we perform comprehensive study of the statistical
properties of the odd harmonic radiation from SASE FEL. The study is
performed in the framework of one-dimensional model with time-dependent
simulation code FAST \cite{fast,book} upgraded for simulation of higher
harmonic generation. We restrict our study with odd harmonics produced
in the SASE FEL. We omit from consideration an effect of
self-consistent amplification of the higher harmonics. In other words,
we solve only electrodynamic problem assuming that particle motion is
governed by the fundamental harmonic. The latter approximation is valid
when power in higher harmonics is much less than in the fundamental.
This does not limit practical applicability of the results: it has been
shown in earlier papers that the growth rate of higher harmonics is too
small to produce visible increase of the coherent amplification above
shot noise in X-ray FELs \cite{kim-1}. Under this approximation and
using similarity techniques we derive universal relations describing
general properties of the odd harmonics in the SASE FEL: power,
statistical and spectral properties. The results are illustrated for
the 3rd and 5th harmonic having practical importance for X-ray FELs.

\section{Basic relations}

The one-dimensional model describes the amplification of the plane
electromagnetic wave by the electron beam in the undulator.  When space
charge and energy spread effects can be neglected, operation of an FEL
amplifier is described in terms of the gain parameter
$\Gamma = \left[
\pi j_0 K_1^2/(I_{A} \lambda_{\mathrm{w}} \gamma^3)
\right]^{1/3}$,
efficiency parameter
$\rho = \lambda_{\mathrm{w}} \Gamma/(4 \pi)$,
and detuning parameter
$\hat{C} = [2\pi/\lambda _ {\mathrm {w}} - \omega (1+K^2/2) /
(2c \gamma^2)]/\Gamma $
(see,e.g. \cite{bon-rho,book}).
Here $\lambda_{\mathrm{w}}$ is undulator period, $K = e
\lambda_{\mathrm{w}} H_{\mathrm{w}} / 2 \pi m c^2$ is undulator
parameter, $\gamma $ is relativistic factor, $H_{\mathrm{w}}$ is
undulator field, $j_0$ is the beam current density,
$(-e)$ and $m$ are
charge and mass of electron, $I_{A} = mc^3/e \simeq 17$~kA, and
$\omega$ is frequency of electromagnetic wave.
Coupling factor $K_h$ is given by

\begin{equation}
K_h = K(-1)^{(h-1)/2}
[J_{(h-1)/2}(Q) - J_{(h+1)/2}(Q)] \ ,
\label{eq:AJJ}
\end{equation}

\noindent $Q = K^2/[2(1+K^2)]$, and $K$ is rms undulator parameter.
When describing start-up from shot noise, one more parameters of the
theory appears -- number of particles in coherence volume,
$N_{\mathrm{c}} = I/(e\rho \omega )$, where $I$ is beam current.

We do not present here general technical details of the time
dependent simulations, they have described previously in details
\cite{fast,book}. The only add-on to these description is particle
loading tool, but it is similar to that described in other papers (see,
e.g. \cite{fawley-loading} and references therein). We note only that
under accepted approximation (particle's dynamics is governed by the
fundamental harmonic) we can simply calculate odd harmonics from
particle distribution, and amplitude of the electric field scales as

\begin{equation}
E(z,t) \propto K_h \int \limits_0^z a_h(z',t-z'/c)dz' \ ,
\label{eq:e-harmonics}
\end{equation}

\noindent where $a_h$ is $h$-th harmonic of the beam bunching. Thus, we
find that coupling factor $K_h$, and time-dependent integral of the
beam bunching become to be factorized. This allows us to extract
universal ratio of the power of higher harmonics to the power of
fundamental harmonic.

\section{Statistical properties of the odd harmonics of the radiation
from SASE FEL}

   In this section we present the results of numerical studies of the
operation of the SASE FEL in the linear and nonlinear regimes. In the
framework of the accepted model, the input parameter of the system is
the number of cooperating electrons $N_{\mathrm{c}}$. Most of the
statistical characteristics of the SASE FEL process are functions of
$N_{\mathrm{c}}$ only in the fixed $z$ coordinate. A typical range of
the values of $N_{\mathrm{c}}$ is $10^6-10^9$ for the SASE FELs of
wavelength range from X-ray up to infrared. The numerical results,
presented in this section, are calculated for the value $N_{\mathrm{c}}
= 3 \times 10^7$ which is typical for a VUV FEL. It is worth mentioning
that the dependence of the output parameters of the SASE FEL on the
value of $N_{\mathrm{c}}$ is rather weak, in fact logarithmic.
Therefore, the obtained results are pretty general and can be used for
the estimation of the parameters of actual devices with sufficient
accuracy.

\subsection{Temporal characteristics}

\begin{figure}[tb]

\includegraphics[width=0.5\textwidth]{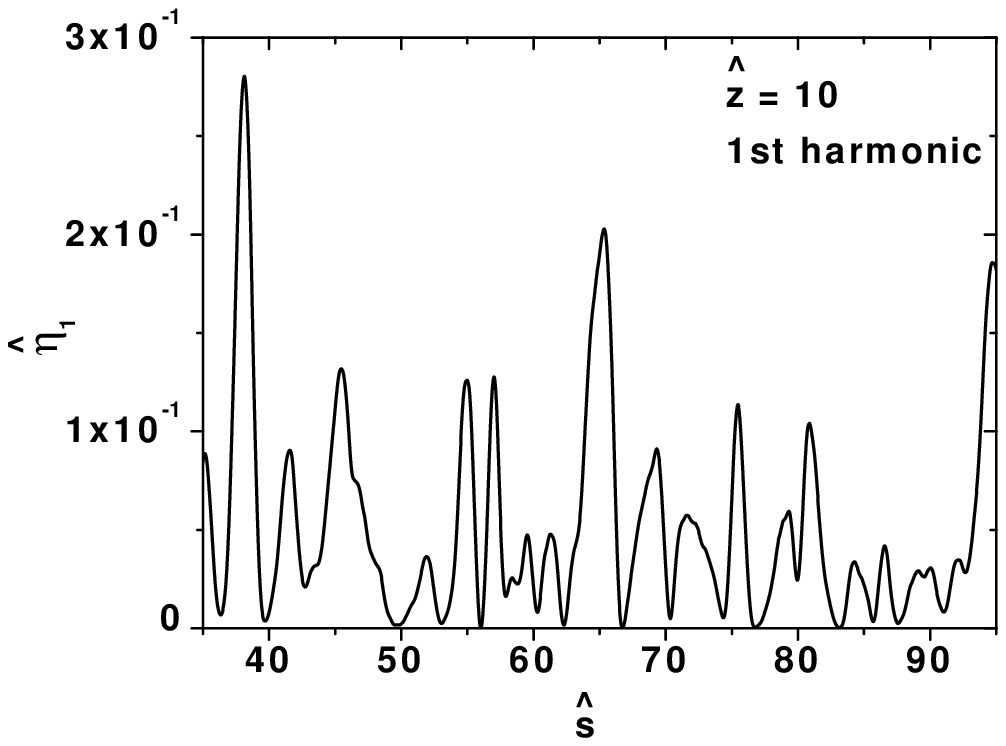}
\includegraphics[width=0.5\textwidth]{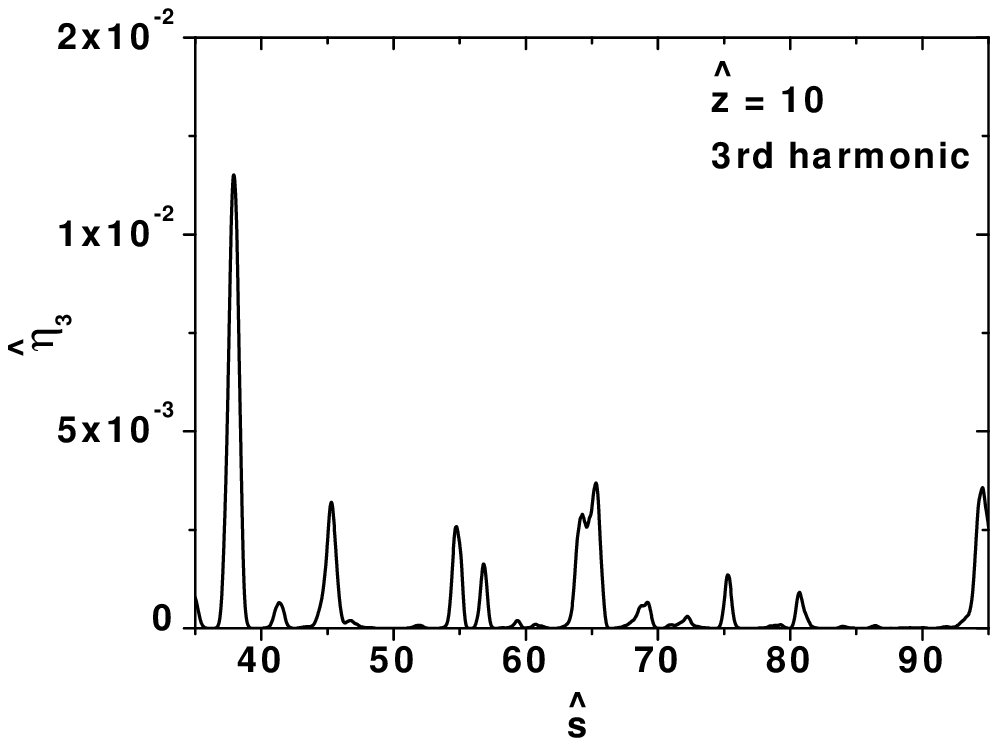}
\vspace*{-15mm}

\includegraphics[width=0.5\textwidth]{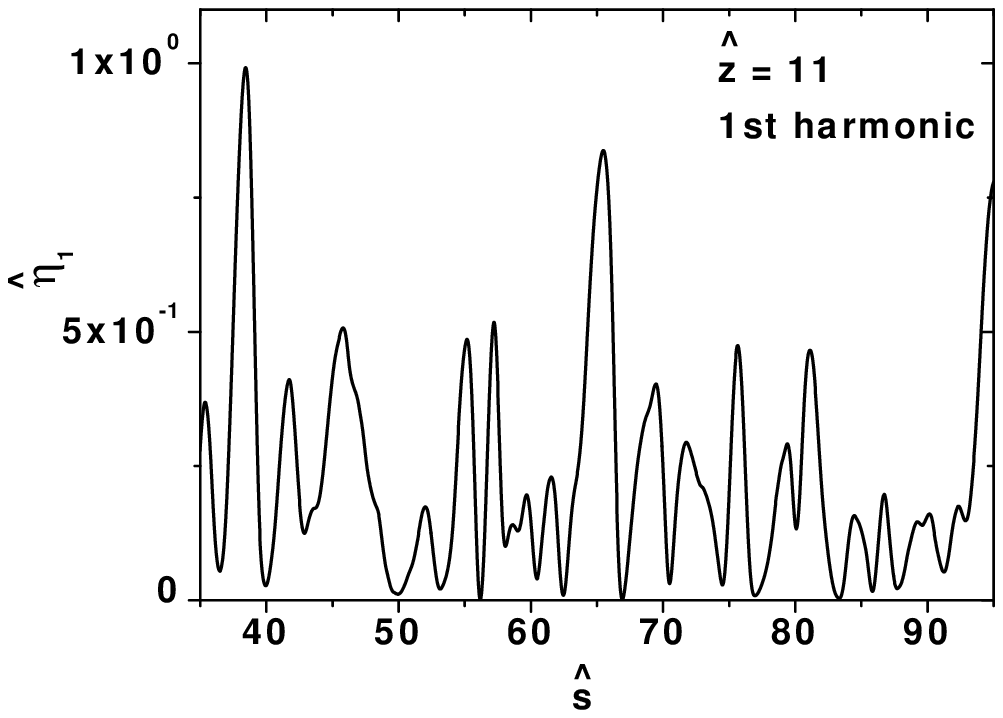}
\includegraphics[width=0.5\textwidth]{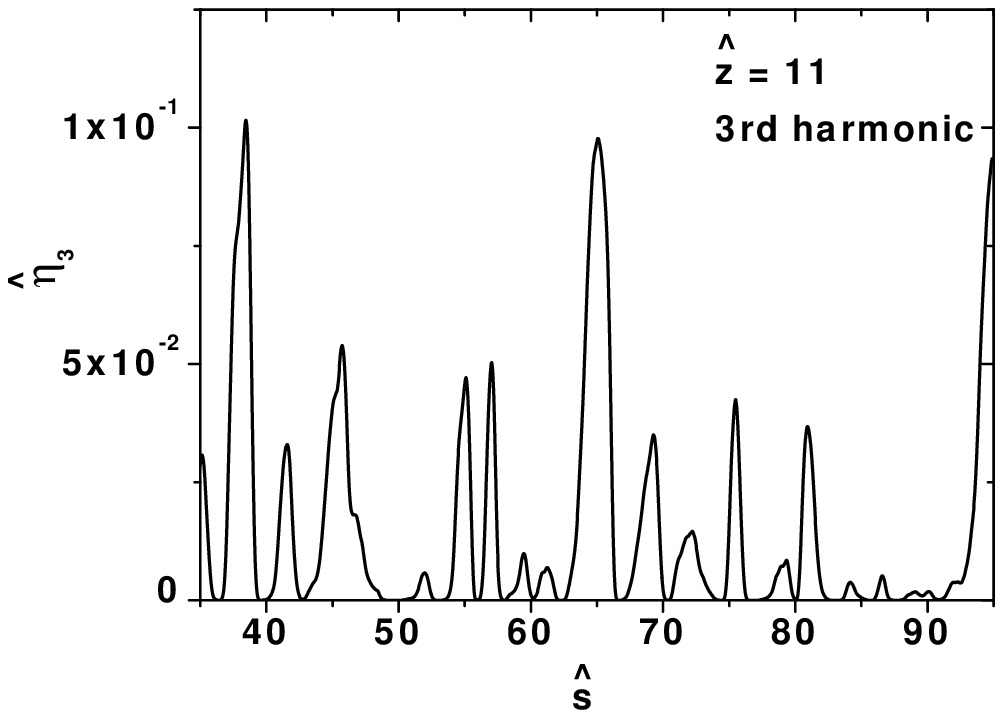}
\vspace*{-15mm}

\includegraphics[width=0.5\textwidth]{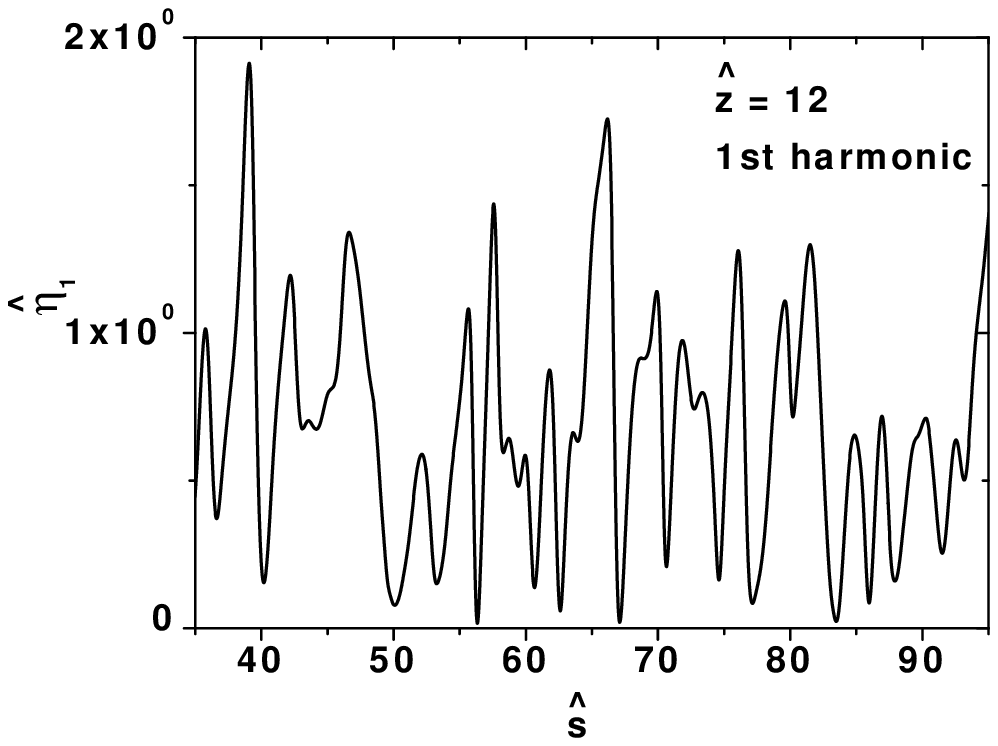}
\includegraphics[width=0.5\textwidth]{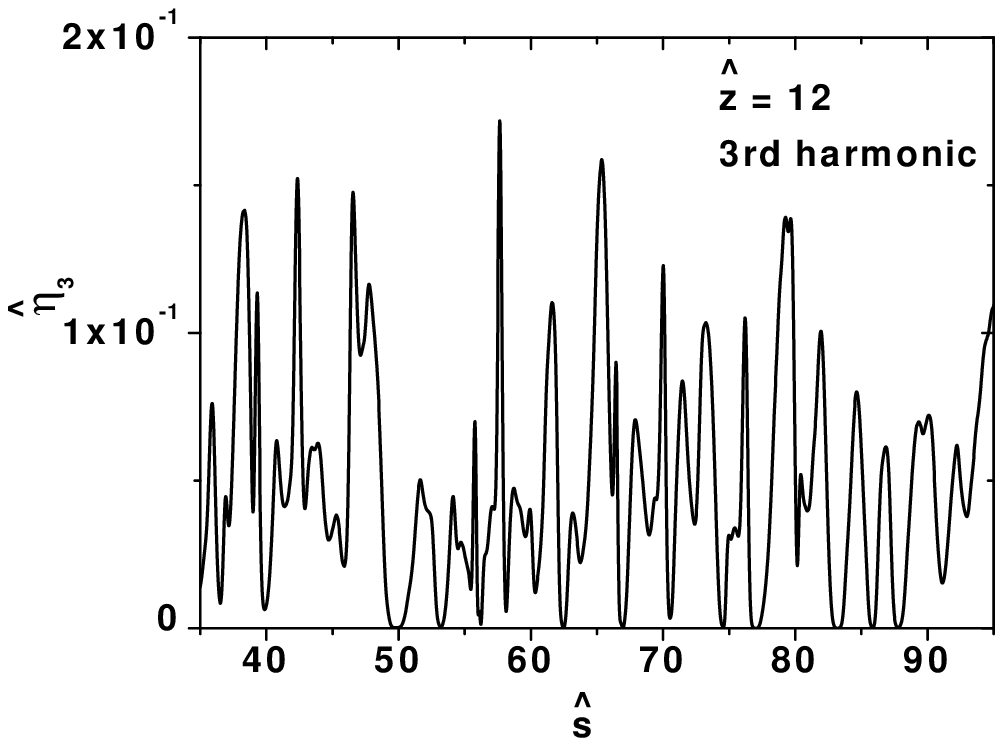}
\vspace*{-15mm}

\includegraphics[width=0.5\textwidth]{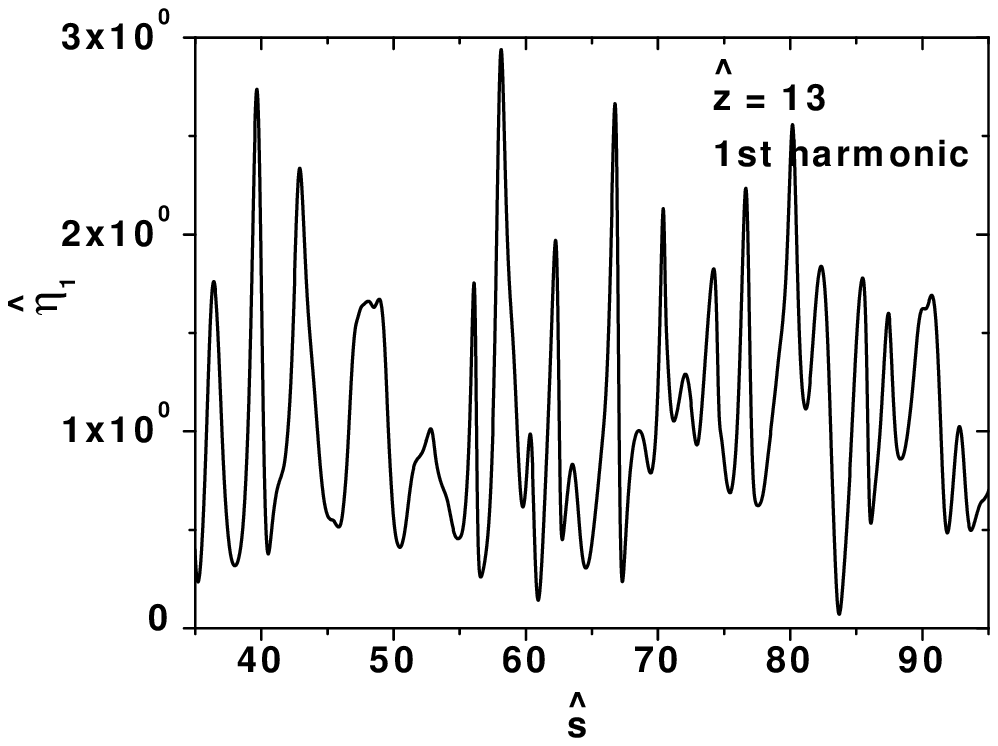}
\includegraphics[width=0.5\textwidth]{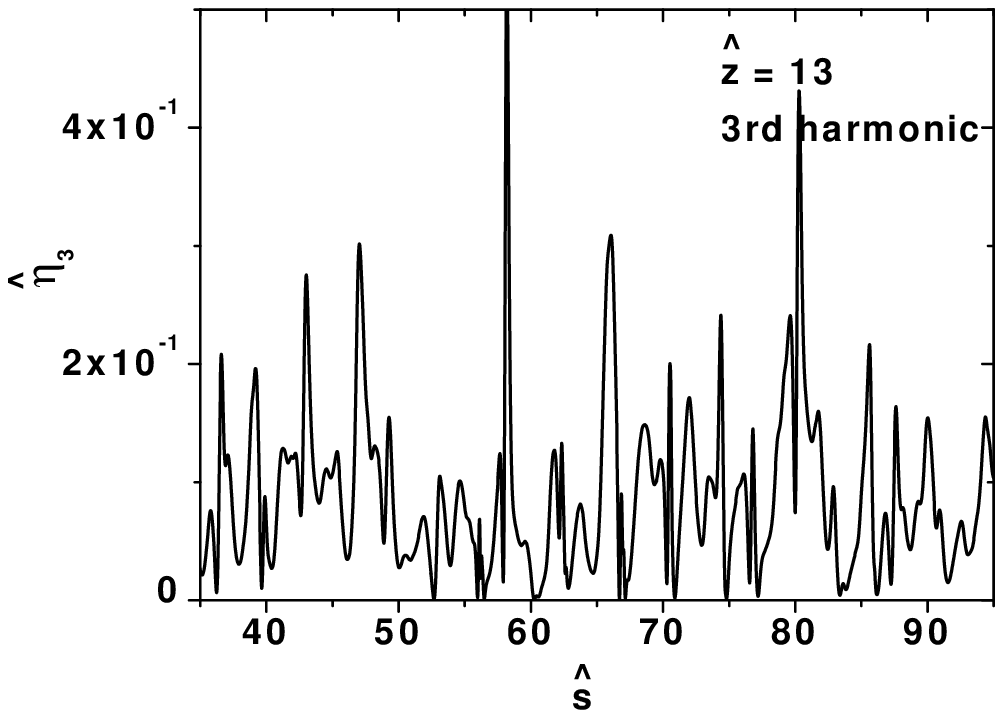}
\vspace*{-5mm}

\caption{
Normalized power in the radiation
pulse versus $\hat{s}=\rho\omega_0(z/\bar{v}_z-t)$ at
different lengths of the FEL amplifier
$\hat{z} = 10-13$.
Left and right columns correspond to the fundamental and 3rd harmonic,
respectively
}

\label{fig:ptemp13}

\end{figure}

\begin{figure}[tb]
\includegraphics[width=0.8\textwidth]{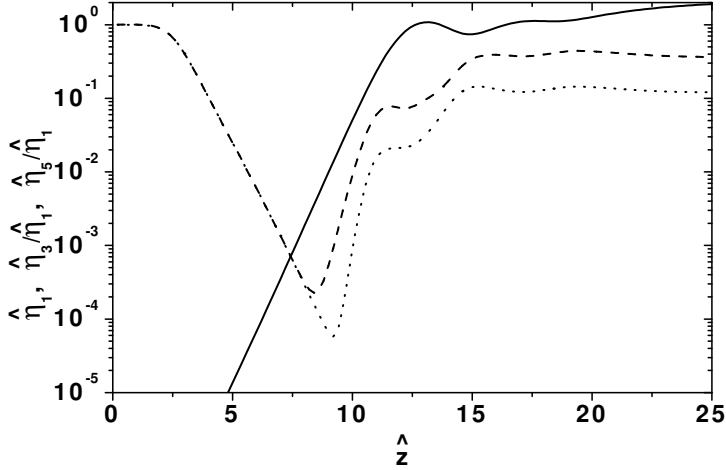}
\caption{
Normalized averaged power of a fundamental harmonic of SASE FEL,
$\hat{\eta }_1 = P_1/(\rho P_{\mathrm{beam}})$, as a function of a
normalized undulator length (solid line). Dashed and dotted lines show a
normalized power ratio, $\hat{\eta }_h/\hat{\eta }_1 = (W_h/W_1)\times
(K_1/K_h)^2$, for the 3rd and 5th harmonic
}
\label{fig:pzharm}
\end{figure}

Figure~\ref{fig:ptemp13} presents a typical time structure of the 1st
and the 3rd harmonic of the radiation from a SASE FEL at different
undulator length $\hat{z} = \Gamma z = 10-13$. Normalized power of
$h$-th harmonic is defined as $\hat{\eta}_h= W_h \times
(K_1/K_h)^2/(\rho W_{\mathrm{b}})$. Longitudinal coordinate along the
pulse is $\hat{s} = \rho \omega_0 (z/\bar{v}_z - t)$.
The head of the pulse is located in the positive direction
of $\hat{s}$. A plot for the averaged power of the 1st harmonic is
shown in Fig~\ref{fig:pzharm} with a solid line. It is seen that
saturation is achieved at the undulator $\hat{z} = 13$. Saturation
length is well described in terms of
the number of cooperating electrons $N_{\mathrm{c}}$
\cite{book,statistics-oc}:

\begin{equation}
\hat{z}_{\mathrm{sat}} \simeq 3 + \frac{1}{\sqrt{3}} \ln N_{\mathrm{c}}
\ .
\label{eq:sat-sase}
\end{equation}

\noindent The normalized efficiency at saturation,
$\hat{\eta}_{\mathrm{sat}} = W_{\mathrm{sat}}/(\rho W_{\mathrm{b}})
\simeq 1.08$, is almost independent of the value of $N_{\mathrm{c}}$.
Dashed and
dotted lines show a normalized power ratio, $\hat{\eta }_h/\hat{\eta
}_1 = (W_h/W_1)\times (K_1/K_h)^2$, for the 3rd and the 5th harmonic. One
can notice that power of the higher harmonics becomes to be above the
shot noise level only in the end of linear regime. This becomes clear
if one takes into account that the shot noise level of the beam
bunching is about $1/\sqrt{N_{\mathrm{c}}}$. We consider an example
typical for VUV FEL with $N_{\mathrm{c}} = 3 \times 10^7$ which
corresponds to the shot noise beam bunching $a \simeq 2 \times 10^{-4}$.
When FEL amplifier operates in the linear regime, odd harmonics grow as
$a_1^h$, and we expect from this simple physical estimation that
coherent contribution into higher harmonics can exceed the shot noise
level only for the values of the beam bunching at the fundamental
harmonic $a_1 \gtrsim 0.1$, i.e. in the end of the linear regime. Note
that shot noise level becomes to be higher when approaching to X-ray
region.

\begin{figure}[tb]
\includegraphics[width=0.8\textwidth]{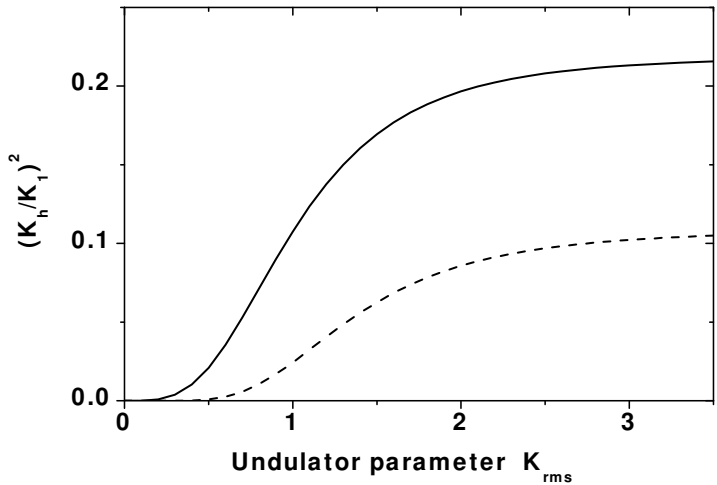}
\caption{
Ratio of coupling factors, $(K_h/K_1)^2$, for the 3rd (solid line) and
the 5th (dashed line) harmonics with respect the fundamental harmonic
versus rms value of undulator parameter $K_{\mathrm{rms}}$
}
\label{fig:ajj135}
\end{figure}

\begin{figure}[tb]
\includegraphics[width=0.8\textwidth]{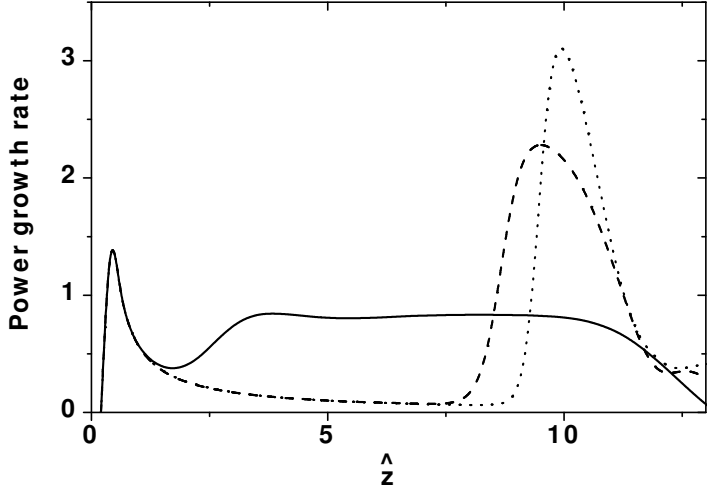}
\caption{
Normalized power growth rate for the 1st, 3rd, and 5th harmonic (solid,
dashed, and dotted line, respectively)
}
\label{fig:pzgrowth}
\end{figure}

The plots presented in Fig.~\ref{fig:ptemp13} allows to trace the
evolution of the 3rd harmonic power from $\hat{z} = 10$ (when it just
started to exceed shot noise level) up to saturation point $\hat{z} =
13$. When the beam bunching at the fundamental harmonic is governed by
a sine function (i.e. SASE FEL operates in the linear regime), we occur
well-known mechanism of the higher harmonic generation, i.e. $a_h
\propto a_1^h$, and spikes of the 3rd harmonic radiation become rather
pronouncing. At this stage, for instance, the growth rate of the higher
harmonics should be proportional to the harmonic number
\cite{hg-3,hg-4,kim-1}. The plot of the power growth rate for the 1st,
3rd, and 5th harmonic is shown in Fig.~\ref{fig:pzgrowth}. One can see
from this plot that in practical situation for SASE FEL the growth
rates of higher harmonics are visibly less than those given by
prediction of the linear theory. This is due to the fact that the value
of $a_1 \simeq 0.1$ can not be considered as a linear stage, and the
beam density modulation is not a sine-like due to nonlinear effects.
The noise nature of the SASE FEL makes a big difference in the behavior
of the growth rates with respect to preidctions given in the framework
of steady-state simulations \cite{hg-3,hg-4}. Analyzing the plot for
the power growth rate we can state that in practical situation
predictions of the steady-state theory are valid only for the 3rd
harmonic, but only on a short piece of undulator close to saturation,
of about one gain length. Quantitative analysis of this stage of
amplification shows that a prediction for the relation between averaged
values of the beam bunching at the third harmonic, $<|a_3|^2> =
6<|a_1|^2>^3$, holds approximately, and is strongly violated for higher
harmonics, because of strong contribution of the shot noise. This
feature of the SASE FEL has been highlighted qualitatively in early
papers \cite{kim-1} with analysis of simulation results obtained with
code GINGER \cite{hg-2a}. Here we just presented more quantitative
study.

The plots in Fig.~\ref{fig:pzharm} present a general result for a ratio
of the power in the higher harmonics with respect to the fundamental
one. For the saturation we find a universal dependency:

\begin{equation}
\frac{\langle W_3 \rangle}{\langle W_1 \rangle} \vert _{\mathrm{sat}} =
0.094 \times \frac{K_3^2}{K_1^2} \ , \qquad
\frac{\langle W_5 \rangle}{\langle W_1 \rangle} \vert _{\mathrm{sat}}=
0.03 \times \frac{K_5^2}{K_1^2} \ .
\label{eq:sat35}
\end{equation}

\noindent Universal functions for the ratio $(K_h/K_1)^2$ are plotted
in Fig.~\ref{fig:ajj135}. Asimptotic values for large
value of undulator parameter are: $(K_3/K_1)^2 \simeq 0.22$, and
$(K_5/K_1)^2 \simeq 0.11$. Thus, we can state that contribution of the
3rd harmonic into the total radiation power of SASE FEL at saturation
could not exceed a level of 2\%. Thus, its influence on the beam
dynamics should be small. This result justifies a basic assumption used
for derivation of a universal relation (\ref{eq:sat35}). A contribution
of the 5th harmonic into the total power at saturation could not exceed
the value of 0.3\%.

\begin{figure}[tb]
\includegraphics[width=0.8\textwidth]{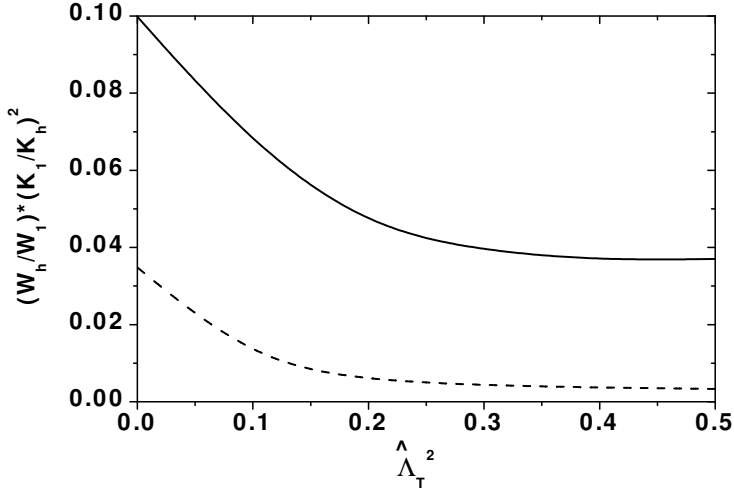}
\caption{
Normalized power ratio at saturation, $(W_h/W_1)\times (K_1/K_h)^2$,
for the 3rd (solid line) and 5th (dashed line) harmonic as a function
of energy spread parameter $\hat{\Lambda }_{\mathrm{T}}^2$. SASE FEL
opeartes at saturation
}
\label{fig:pharmesp}
\end{figure}

Another important topic is an impact of the electron beam quality on
the nonlinear harmonic generation process. In the framework of the
one-dimensional theory this effect is described with the energy spread
parameter $\hat{\Lambda }_{\mathrm{T}}^2$ \cite{book}:

\begin{displaymath}
\hat{\Lambda}^2_{\mathrm{T}} =
\frac{\langle (\Delta {E})^2 \rangle }{\rho^2 {E}_{0}^2} \ ,
\end{displaymath}

\noindent where $\langle (\Delta {E})^{2}\rangle $ is the rms energy
spread. Thus, result given by (\ref{eq:sat35}) is generalized to the
case of finite energy spread with the plot presented in
Fig.~\ref{fig:pharmesp}.
We
see that the energy spread in the electron beam suppresses power of the
higher harmonics. Within practical range of $\hat{\Lambda
}_{\mathrm{T}}^2$ this suppression can be about a factor of 3 for the
3rd harmonic, and about an order of magnitude for the 5th harmonic.
For practical estimations one should use an effective value of the
energy spread describing contribution of the energy spread and
emiitance to the longitudinal velocity spread \cite{book}:

\begin{displaymath}
\frac{\langle (\Delta {E})^2 \rangle_{\mathrm{eff}}}{{E}_{0}^2}
=
\frac{\langle (\Delta {E})^2 \rangle}{{E}_{0}^2} +
\frac{2\gamma_{z}^4 \epsilon^2}{ \beta ^2} \ ,
\end{displaymath}

\noindent where $\gamma_{z}$ is longitudinal relativistic factor,
$\epsilon $ is beam emittance, and $\beta $ is focusing beta-function.
The plot in Fig.~\ref{fig:pharmesp} covers practical range of
parameters fo X-ray FELs. The saturation length at $\hat{\Lambda
}_{\mathrm{T}}^2 = 0.5$ is increased by a factor of 1.5 with respect to
the "cold" beam case $\hat{\Lambda }_{\mathrm{T}}^2 = 0$.

\subsection{Probability distributions}

\begin{figure}[b]
\includegraphics[width=0.8\textwidth]{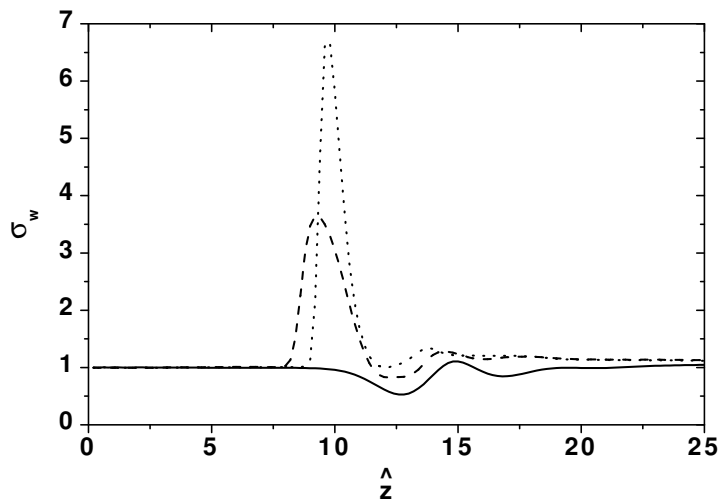}
\caption{
Normalized rms deviation of the fluctuations of the instanteneous
radiation power as a function of the normalized undulator length.
Solid, dashed, and
dotted lines correspond to the fundamental, 3rd, and 5th harmonic,
respectively
}
\label{fig:sigpharm}
\end{figure}

The next step of our study is the behavior of the probability
distribution of the instantaneous power. In Fig.~\ref{fig:sigpharm} we
show the normalized rms deviation of the instantaneous radiation power,
$\sigma _{\mathrm{w}} = \langle (W-\langle W \rangle )^2 \rangle
^{1/2}/ \langle W \rangle $, as a function of the undulator length. We
see that at the initial stage of SASE FEL operation rms deviation of
the instantaneous power is equal to one for all harmonics. This is a
consequence of start-up from the shot noise in the electron beam:
statistical properties of the undulator radiation and of the radiation
from SASE FEL operating in the linear regime are governed by Gaussian
statistics \cite{book,statistics-oc}. One of the important features of
the Gaussian statistics is that the normalized rms deviation of the
instantaneous radiation power is equal to the unity. For the
fundamental harmonic statistics of the radiation becomes to be
non-Gaussian when the amplification process enters non-linear mode
\cite{book,statistics-oc}. For the higher harmonics non-Gaussian
statistics takes place when the nonlinear harmonic generation starts to
dominate above incoherent radiation (at $\hat{z} \gtrsim 8$ in the
present numerical example). Analytical theory of nonlinear harmonic
generation \cite{kim-1} predicts the value of $\sigma _{\mathrm{w}}
\simeq 4$ for the third harmonic. Analysis of the relevant curve in
Fig.~\ref{fig:sigpharm} shows that this prediction holds approximately
in a short piece of the undulator length only. As we explained above,
this is due to the fact that nonlinear harmonic generation starts to
dominate above incoherent radiation only at the values of the beam
bunching at the fundamental harmonic $a_1 \sim 0.1$. However, at such a
value of the beam bunching the modulation of the beam density already
deviates from a sin-like shape due to nonlinear effects.

\begin{figure}[tb]

\includegraphics[width=0.5\textwidth]{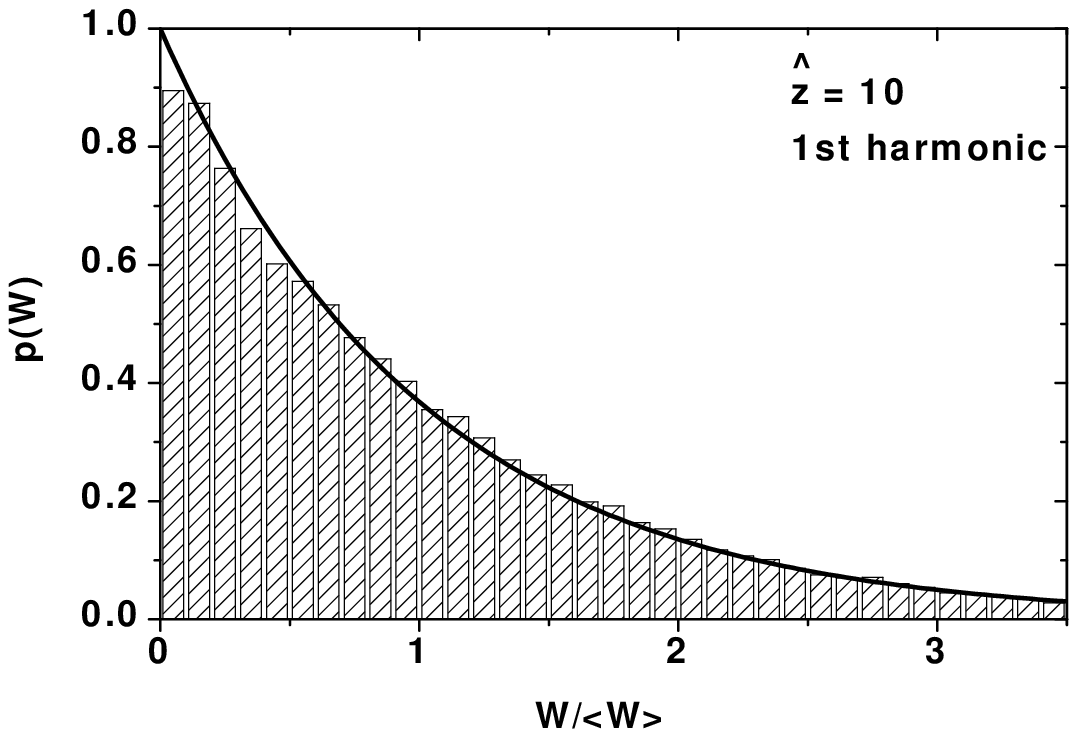}
\includegraphics[width=0.5\textwidth]{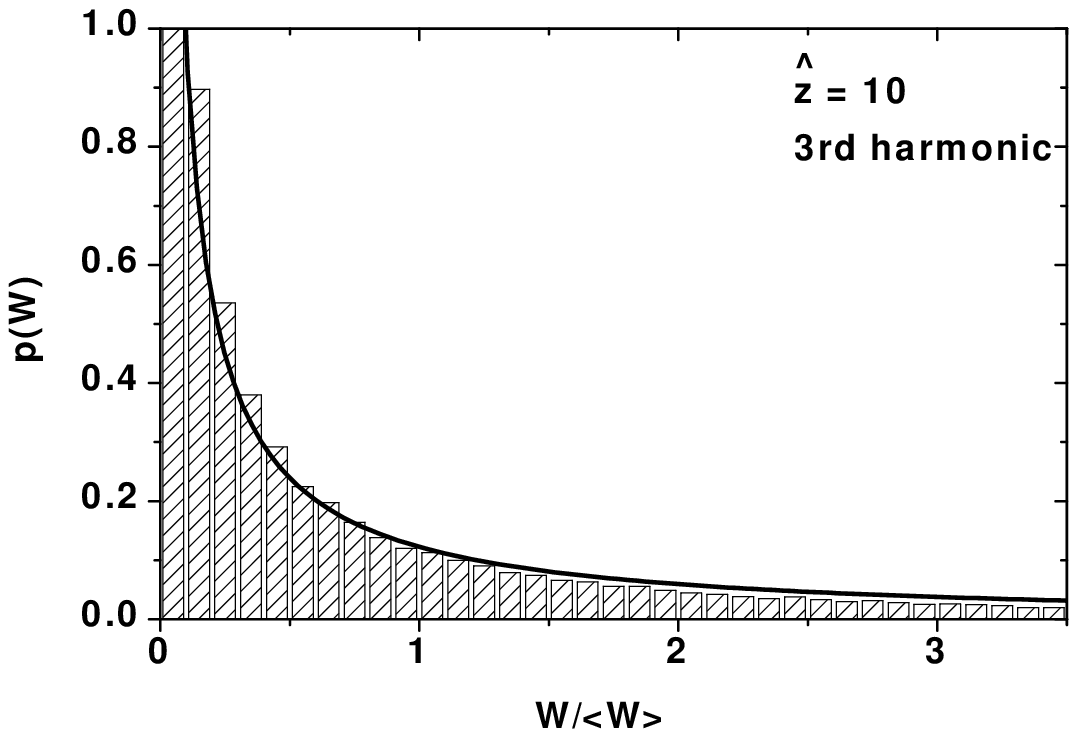}
\vspace*{-15mm}

\includegraphics[width=0.5\textwidth]{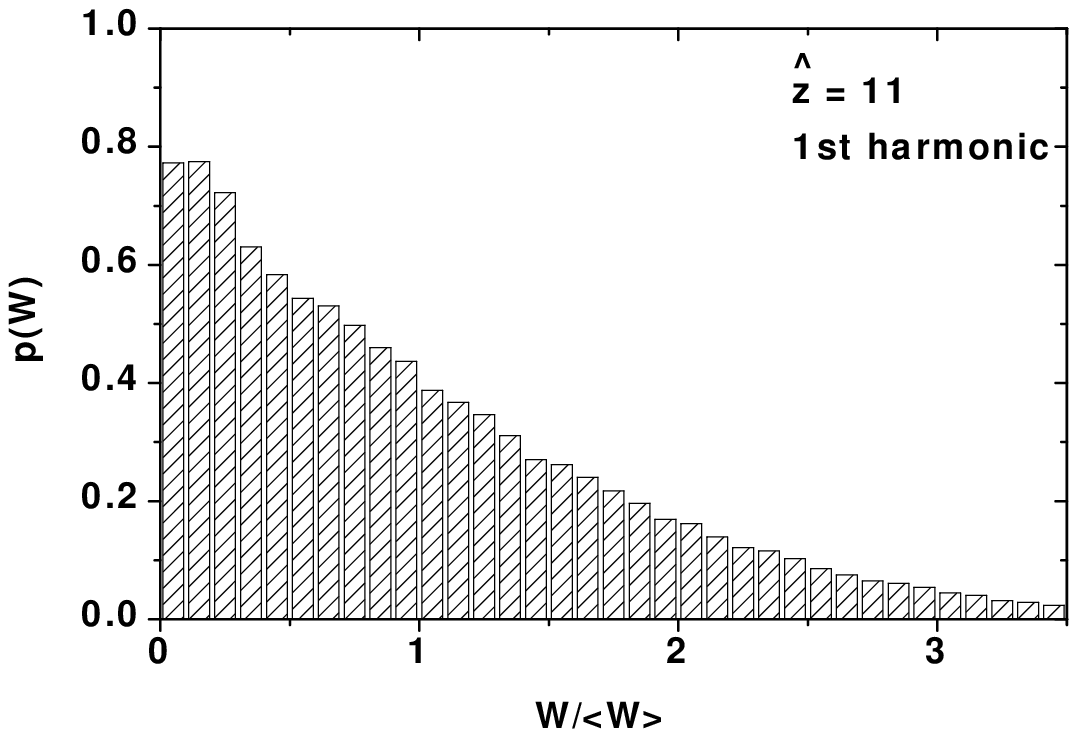}
\includegraphics[width=0.5\textwidth]{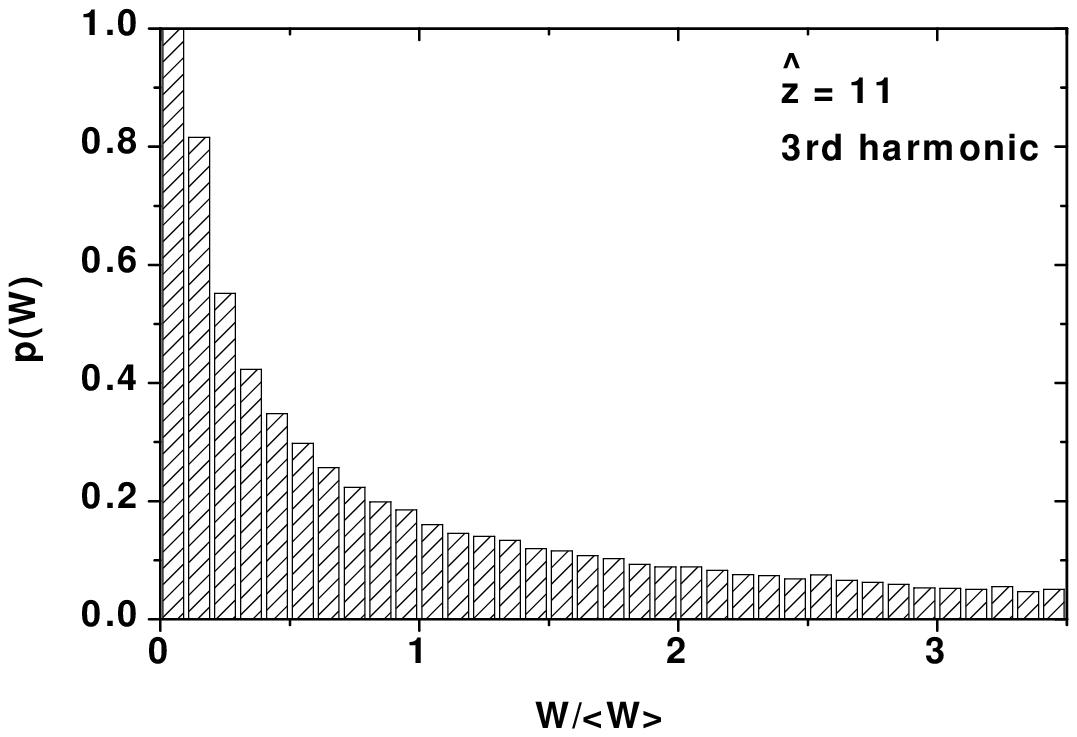}
\vspace*{-15mm}

\includegraphics[width=0.5\textwidth]{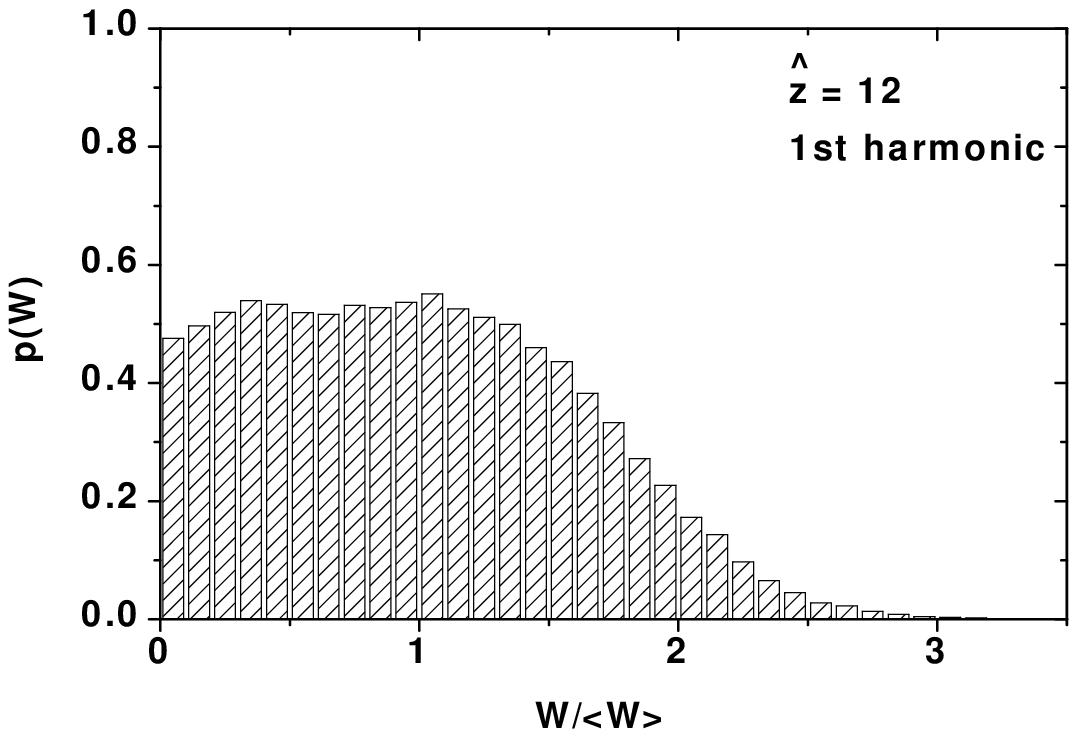}
\includegraphics[width=0.5\textwidth]{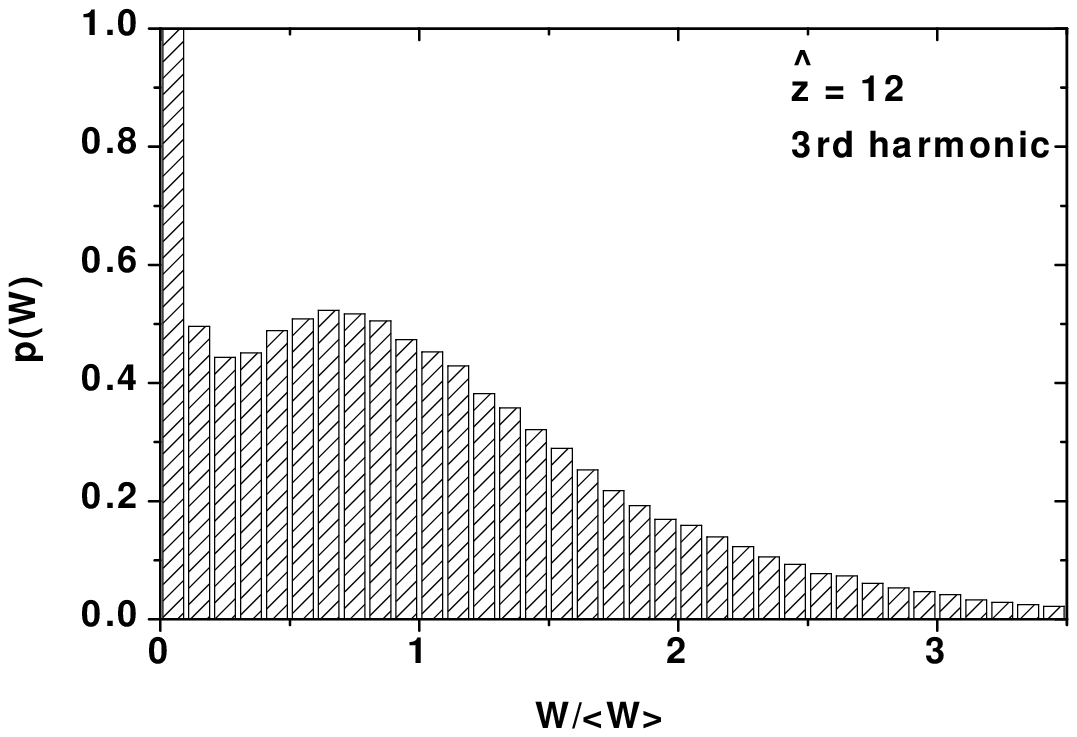}
\vspace*{-15mm}

\includegraphics[width=0.5\textwidth]{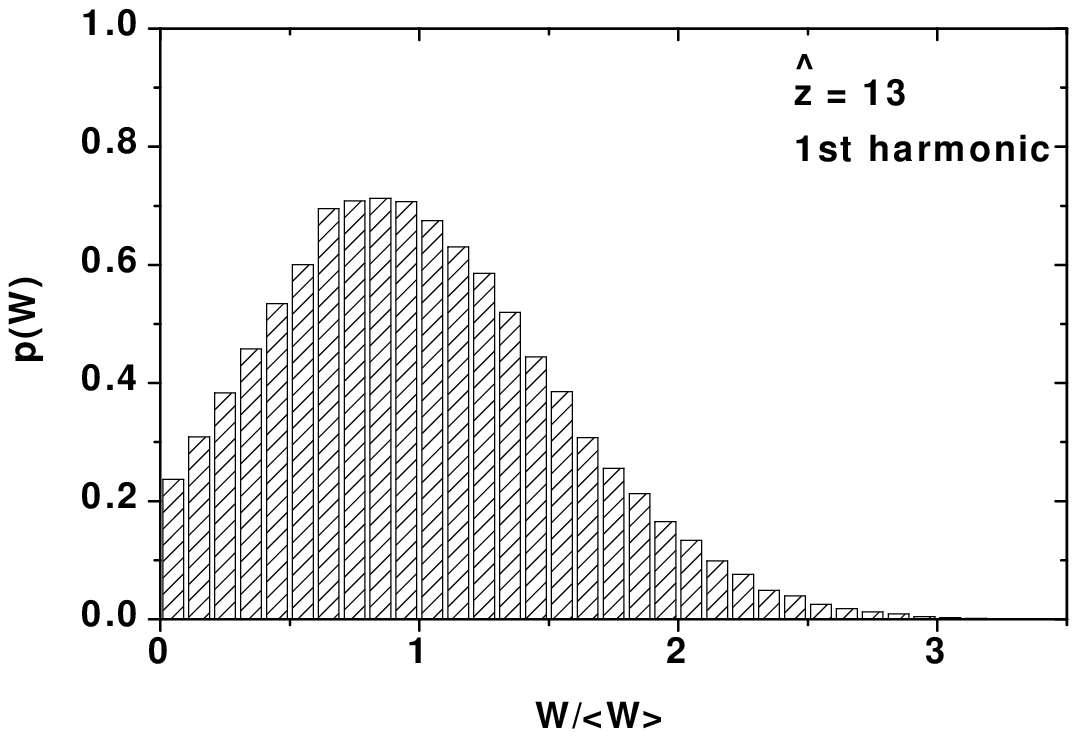}
\includegraphics[width=0.5\textwidth]{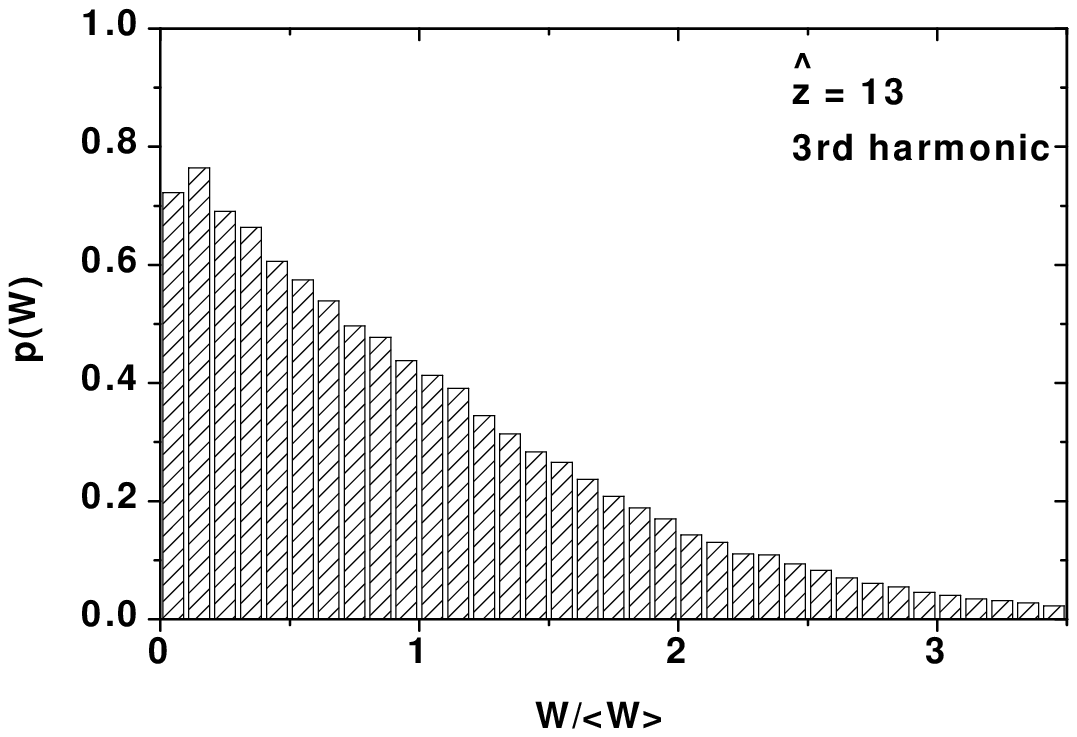}
\vspace*{-5mm}

\caption{
Probability distribution of instanteneous radiation power
at different lengths of the FEL amplifier $\hat{z} = 10-13$.
Left and right columns correspond to the fundamental and 3rd harmonic,
respectively. Solid line shows probability density function
(\ref{eq:prob-hg})
}

\label{fig:hist135}

\end{figure}

Probability density distributions for the instantaneous power of the
fundamental and the 3rd harmonic are presented in
Fig.~\ref{fig:hist135}. SASE radiation is a stochastic object and at a
given time it is impossible to predict the amount of energy which flows
to a detector.  The initial modulation of the electron beam is defined
by the shot noise and has a white spectrum. The high-gain FEL amplifier
cuts and amplifies only a narrow frequency band of the initial spectrum
$\Delta\omega/\omega \ll 1$. In the time domain, the temporal structure
of the fundamental harmonic radiation is chaotic with many random
spikes, with a typical duration given by the inverse width of the
spectrum envelope.  Even without performing numerical simulations, we
can describe some general properties of the fundamental harmonic of the
radiation from the SASE FEL operating in the linear regime. Indeed, in
this case we deal with Gaussian statistics. As a result, the
probability distribution of the instantaneous radiation intensity $W$
should be the negative exponential probability density distribution
\cite{book,statistics-oc}:

\begin{equation}
p(W) = \frac{1}
{\langle W \rangle }
\exp\left(-\frac{W}
{\langle W \rangle }\right) \ .
\label{neg-exp-3}
\end{equation}

Here one should realize clearly that the notion of instantaneous
intensity refers to a certain moment in time, and that the analysis
must be performed over an ensemble of pulses. Also, the energy in the
radiation pulse $E$ should fluctuate in accordance with the gamma
distribution \cite{book,statistics-oc}:

\begin{equation}
p(E) =
\frac{M^{M}}{\Gamma(M)}\left(\frac{E}{\langle{E}\rangle}\right)^{M-1}
\frac{1}{\langle{E}\rangle}
\exp\left(-M\frac{E}{\langle{E}\rangle}\right) \ ,
\label{eq:gamma-distribution}
\end{equation}

\noindent where $\Gamma(M)$ is the gamma function of argument $M$, and
$1/M = \langle{(E-\langle{E}\rangle)^{2}\rangle}/\langle{E}\rangle^{2}$
is the normalized dispersion of the energy distribution. These
properties are well known in statistical optics as properties of
completely chaotic polarized radiation \cite{goodman}.

The statistics of the high-harmonic radiation from the SASE FEL changes
significantly with respect to the fundamental harmonic (e.g., with
respect to Gaussian statistics).  It is interesting in our case to be
able to determine the probability density function of instantaneous
intensity of SASE radiation  after it has been subjected to nonlinear
transformation. We know the probability density function $p(W) =
\langle{W}\rangle^{-1}\exp(-W/\langle{W}\rangle)$ of the fundamental
intensity $W$, and $W$ is subjected to a transformation $z = (W)^{n}$.
The problem is then to find the probability density function $p(z)$. It
can be readily shown that this probabilty distribution is
\cite{attofel}:

\begin{equation}
p(z) =
\frac{z}{n\langle{W}\rangle}
{z}^{(1-n)/n}\exp(-z^{1/n}/\langle{W}\rangle)
\ .
\label{eq:prob-hg}
\end{equation}

\noindent Using this distribution we get the expression for the mean
value: $\langle{z}\rangle = n!\langle{W}\rangle^{n}$. Thus, the
$n$th-harmonic radiation for the SASE FEL has an intensity level
roughly $n!$ times larger than the corresponding steady-state case, but
with more shot-to-shot fluctuations compared to the fundamental
\cite{kim-1}. Nontrivial behavior of the intensity of the
high harmonic reflects the complicated nonlinear transformation of the
fundamental harmonic statistics. One can see that Gaussian statistics
is no longer valid. Upper plots in Fig.~\ref{fig:hist135} give an
illustration to these consideration. Despite in practical example we do
not have pure linear amplification regime, the probability density
functions for the instanteneous power follow rather well prediction
(\ref{eq:prob-hg}).

Analysis of the probability distributions in Fig.~\ref{fig:hist135}
shows that in the nonlinear regime, near the saturation point, the
distributions change significantly with respect to the linear regime for
both, the fundamental and the 3rd harmonic. An important message is
that at the saturation point the 3rd harmonic radiation exhibits much
more noisy behaviour (nearly negative exponential) while stabilization
of the fluctuations of the fundamental harmonics takes place.

\subsection{Correlation functions}

The first and the second order time correlation functions are defined
follows:

\begin{eqnarray}
g_1(t-t') & = &
\frac{\langle \tilde{E}(t)\tilde{E}^*(t')\rangle }{\left[\langle |\tilde{E}(t)|^2\rangle
\langle |\tilde{E}(t')|^2\rangle \right]^{1/2}} \ ,
\nonumber \\
g_2(t-t')  & = & \frac{\langle |\tilde{E}(t)|^2|\tilde{E}(t')|^2\rangle
} {\langle |\tilde{E}(t)|^2\rangle \langle |\tilde{E}(t')|^2\rangle }
\ . \label{def-corfunction} \end{eqnarray}

\begin{figure}[tb]

\includegraphics[width=0.5\textwidth]{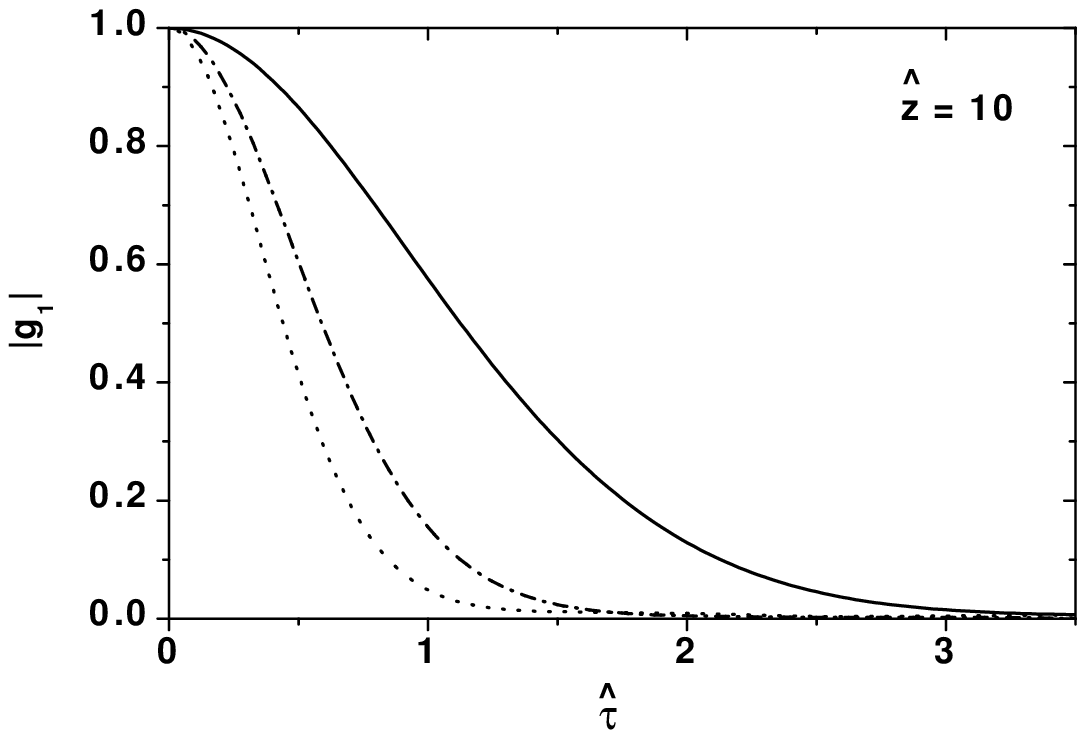}
\includegraphics[width=0.5\textwidth]{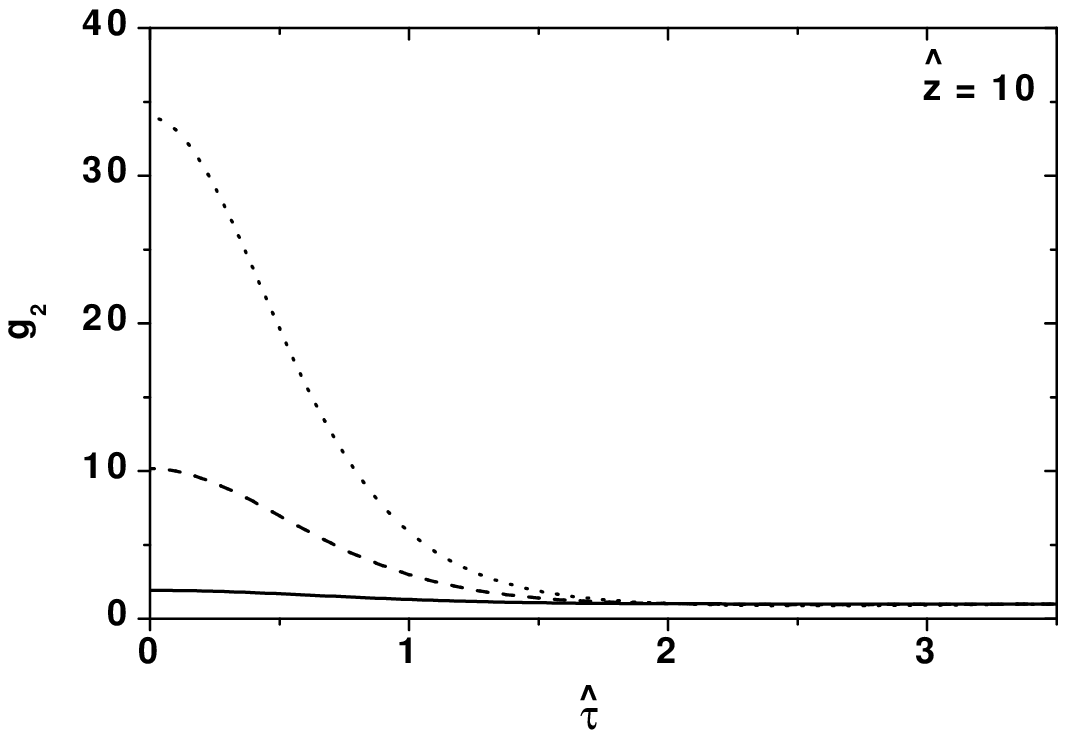}
\vspace*{-15mm}

\includegraphics[width=0.5\textwidth]{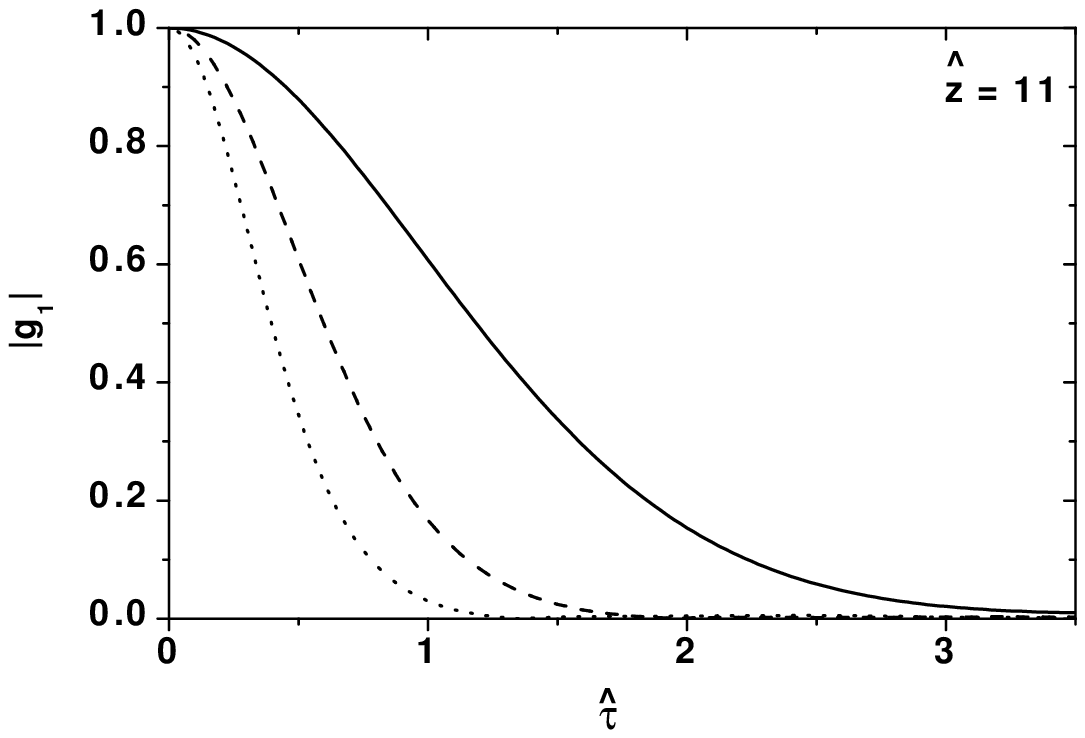}
\includegraphics[width=0.5\textwidth]{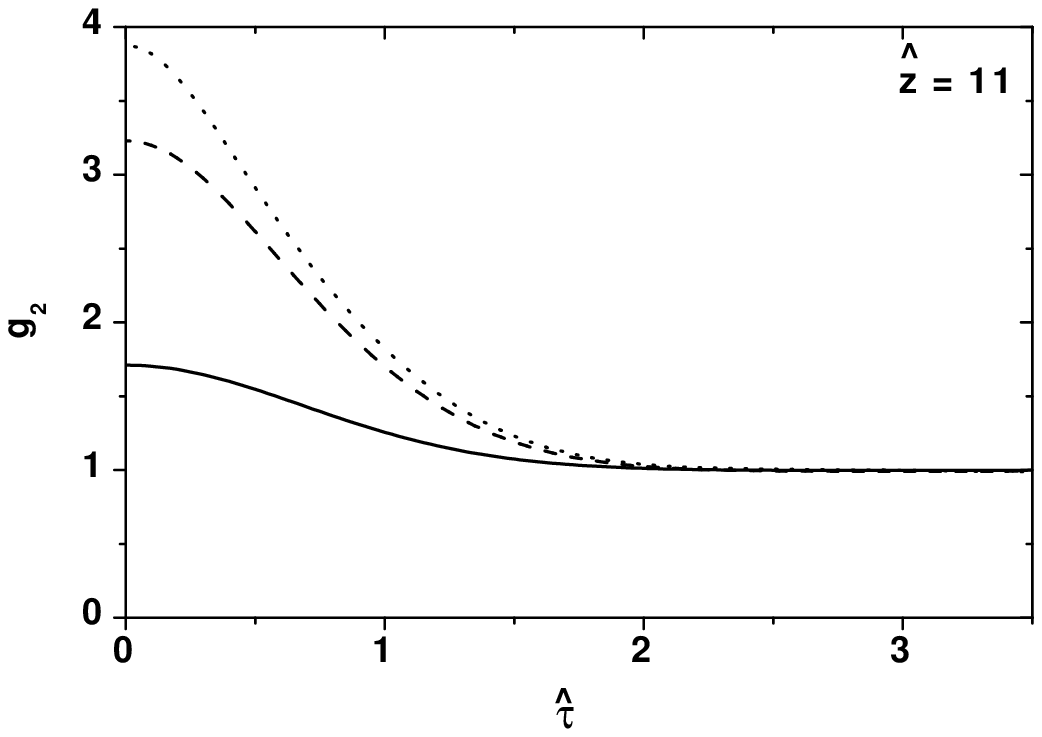}
\vspace*{-15mm}

\includegraphics[width=0.5\textwidth]{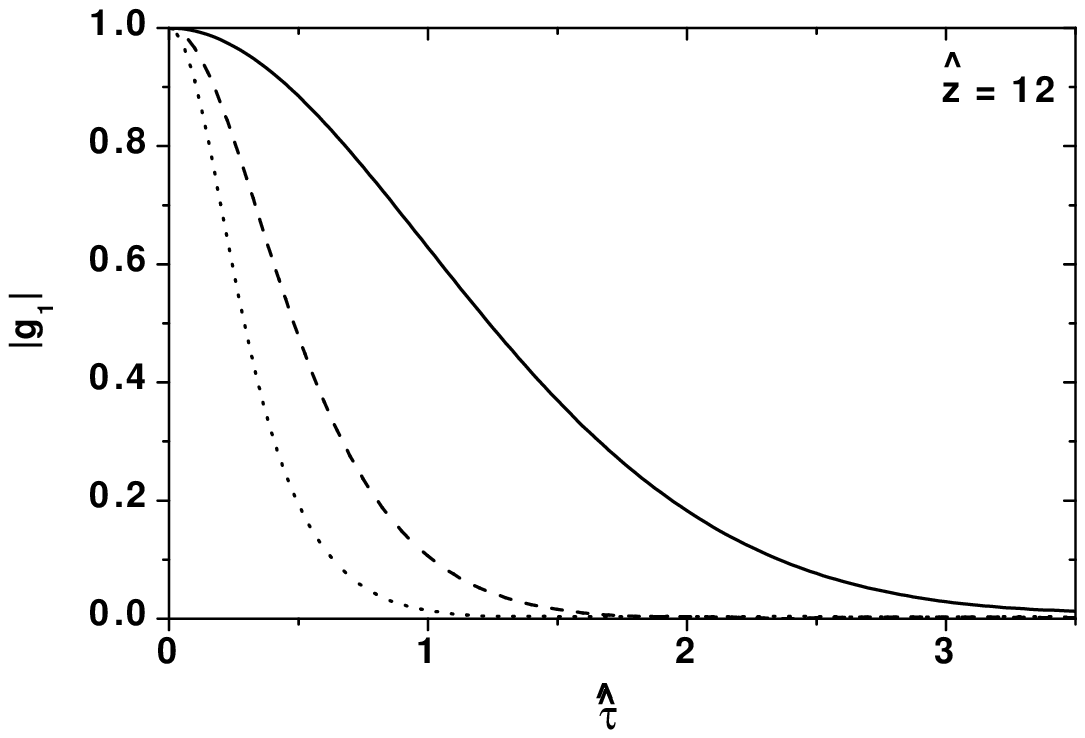}
\includegraphics[width=0.5\textwidth]{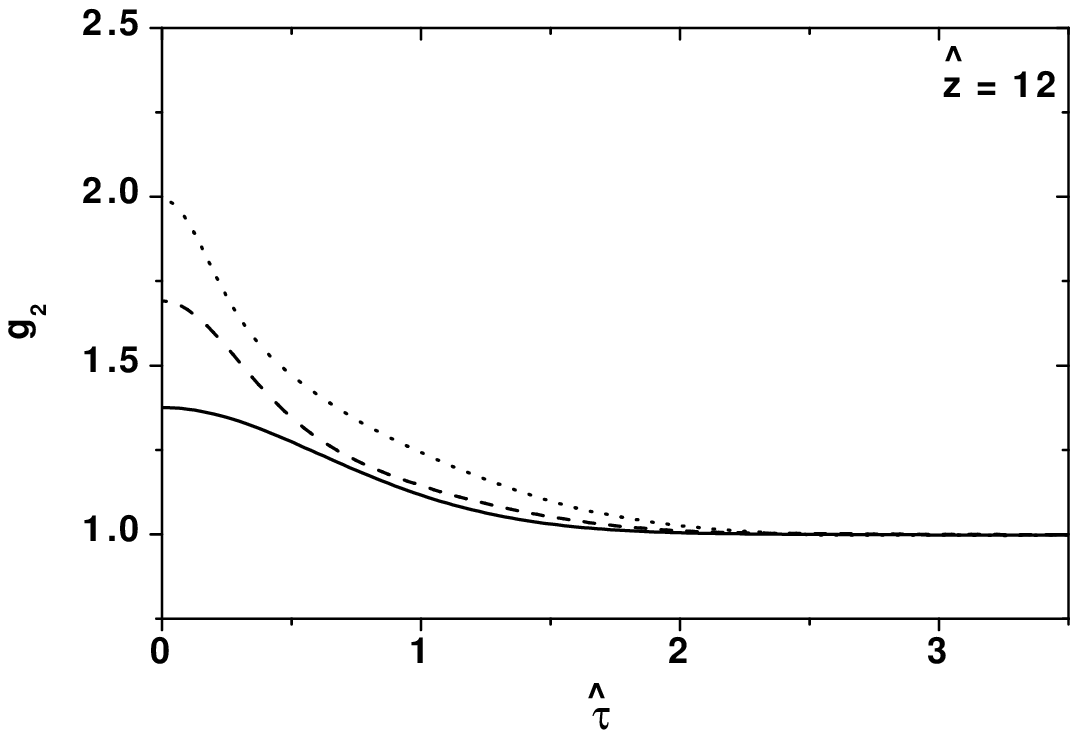}
\vspace*{-15mm}

\includegraphics[width=0.5\textwidth]{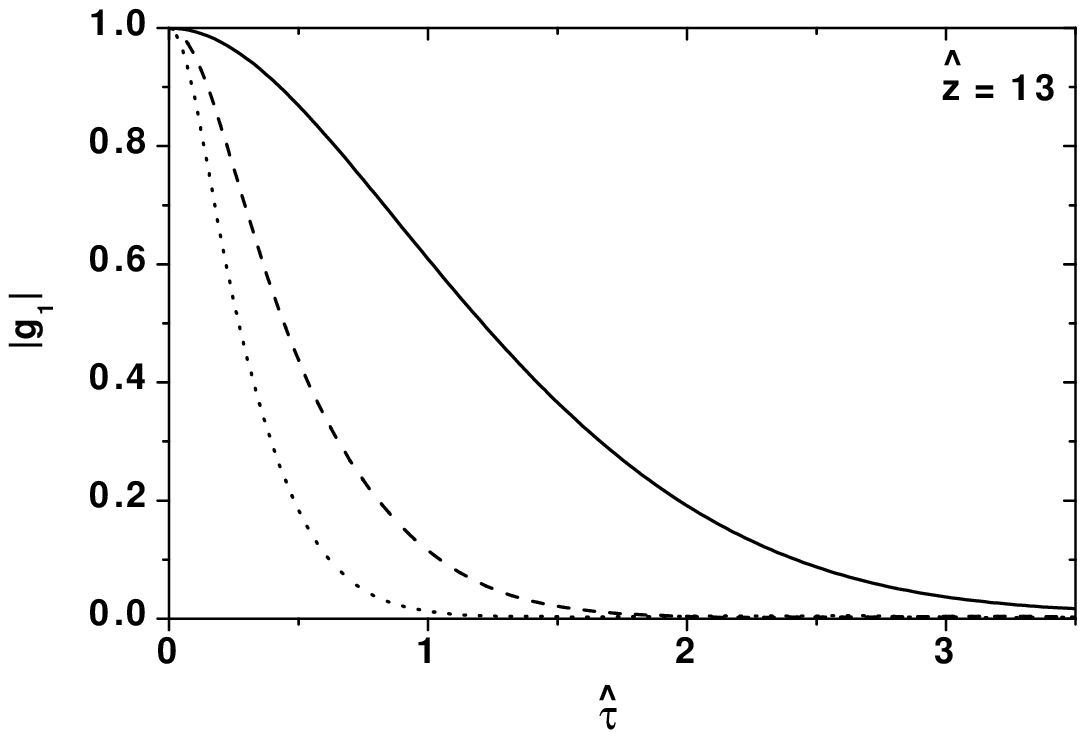}
\includegraphics[width=0.5\textwidth]{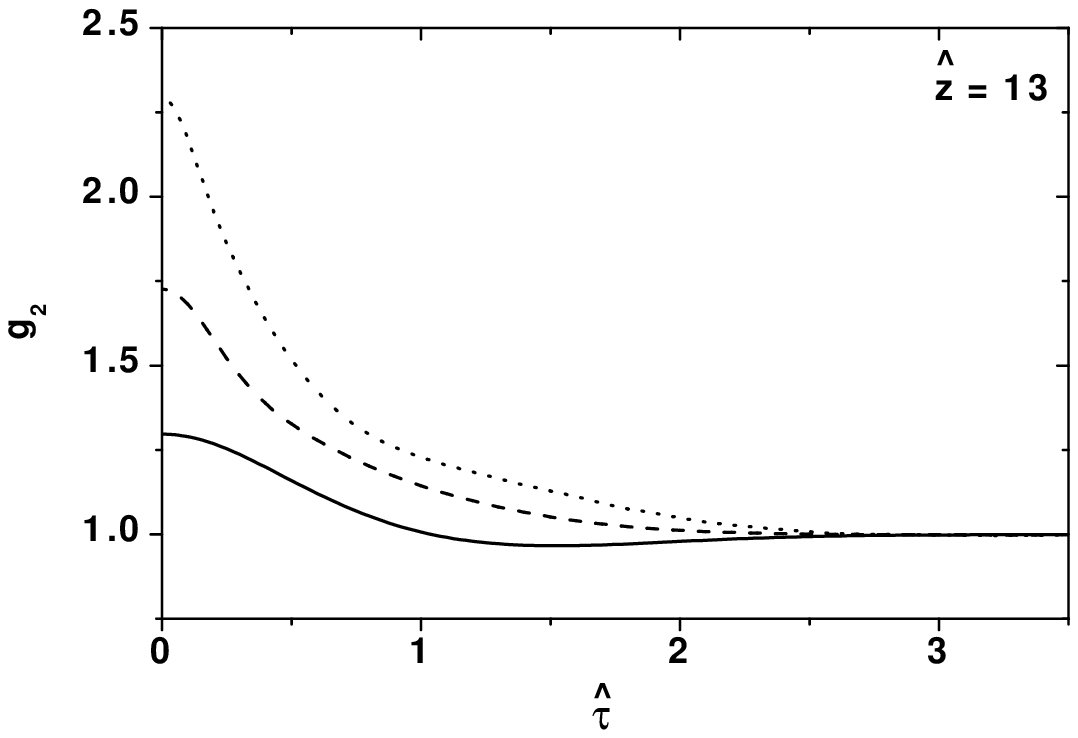}
\vspace*{-5mm}

\caption{
First (left column) and second (right column) order correlation
function at
different lengths of the FEL amplifier
$\hat{z} = 10-13$.
Solid, dashed, and dotted lines correspond to the fundamental,
3rd and 5th harmonic, respectively
}

\label{fig:acor135}

\end{figure}

    In Fig.~\ref{fig:acor135} we show the evolution of
the time correlation functions of first and second order. At
each normalized position along the undulator, $\hat{z}$, they are
plotted versus the normalized variable
$\hat{\tau} =  \rho \omega_0 (t - t')$.
Upper plot in Fig.~\ref{fig:acor135} corresponds to the linear stage of
SASE FEL operation. In the case of the fundamental harmonice we
deal with a Gaussian random process and the relation
between the correlation functions holds for $g_2(t-t') = 1 + |
g_1(t-t')|^2$. This feature does not hold place for figher harmonics.
The nontrivial behavior of the correlation functions reflects the
complicated nonlinear evolution of the SASE FEL process.
The second-order correlation
function of zero argument, $g_2(0)$, takes values smaller or larger
than two, but always larger than unity. Note that there is a simple
relation between $g_2(0)$ and the normalized rms power deviation:
$g_2(0) = 1 + \sigma_{\mathrm{w}}^2$ (see Fig.~\ref{fig:sigpharm}). It
is a well-known result of statistical optics that the cases of
$g_2(0)=1$ and $g_2(0)=2$ correspond to stabilized single-mode laser
radiation and to completely chaotic radiation from a thermal source,
respectively. The values of $g_2(0)$ between 1 and 2 belong to some
intermediate situation. In classical optics, a radiation source with
$g_2(0) < 1$ cannot exist but the case of $g_2(0) > 2$ is possible.  As
one can see from Fig.~\ref{fig:acor135}, the latter
phenomenon (known as superbunching) occurs for higher harmonics of SASE
FEL, or for fundamental one when the SASE FEL operating in the
nonlinear regime.

\begin{figure}[tb]
\includegraphics[width=0.8\textwidth]{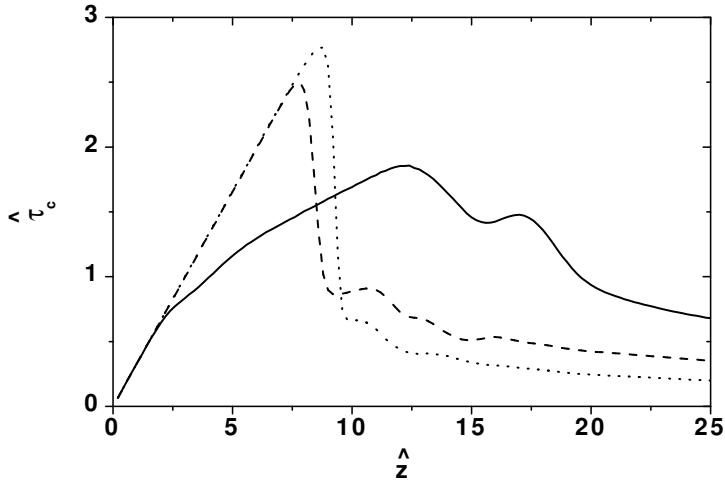}
\caption{
Normalized coherence time of a SASE FEL as a function of normalized
undulator length.
Solid, dashed, and
dotted lines correspond to the fundamental, 3rd, and 5th harmonic,
respectively
}
\label{fig:tcor135}
\end{figure}

    In Fig.~\ref{fig:tcor135} we present the dependence on the undulator
length of the normalized coherence time $\hat{\tau}_{\mathrm{c}} = \rho
\omega_0 \tau_{\mathrm{c}}$, where $\tau_{\mathrm{c}}$ is

\begin{equation}
\tau_{\mathrm{c}} = \int \limits^{\infty}_{-\infty}
| g_1(\tau) |^2 \D\tau \ .
\label{coherence-time-def}
\end{equation}

For the fundamental harmonic the coherence time
achieves its maximal value near the saturation point and then decreases
drastically. The maximal value of $\hat{\tau}_{\mathrm{c}}$ depends on
the saturation length and, therefore, on the value of the parameter
$N_{\mathrm{c}}$. With logarithmic accuracy we haave the following
expression for the coherence time of the fundamental harmonic:

\begin{displaymath}
(\hat{\tau}_{\mathrm{c}})_{\mathrm{max}} \simeq
\sqrt{ \frac{\pi \ln N_{\mathrm{c}} }{18} } \ .
\end{displaymath}

\noindent One can find from Fig.~\ref{fig:acor135} that coherence time
at saturation for higher harmonics approximately falls inversely
proprtional to the harmonic number $h$.

\subsubsection{Spectral Characteristics}

\begin{figure}[b]

\includegraphics[width=0.5\textwidth]{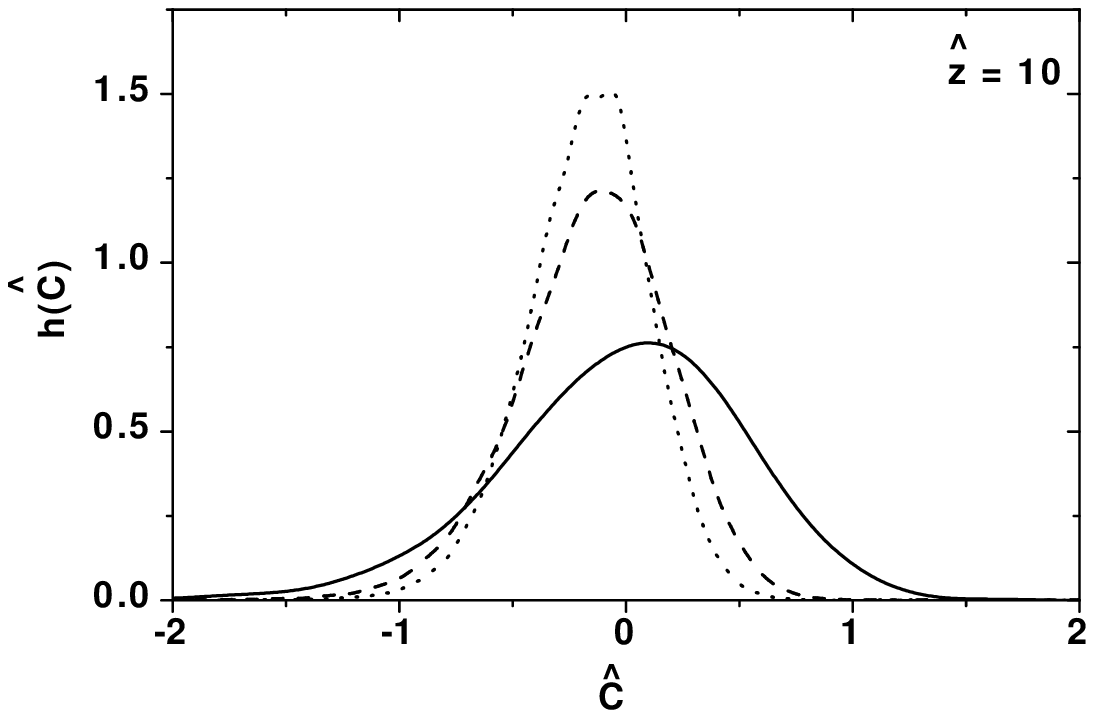}
\includegraphics[width=0.5\textwidth]{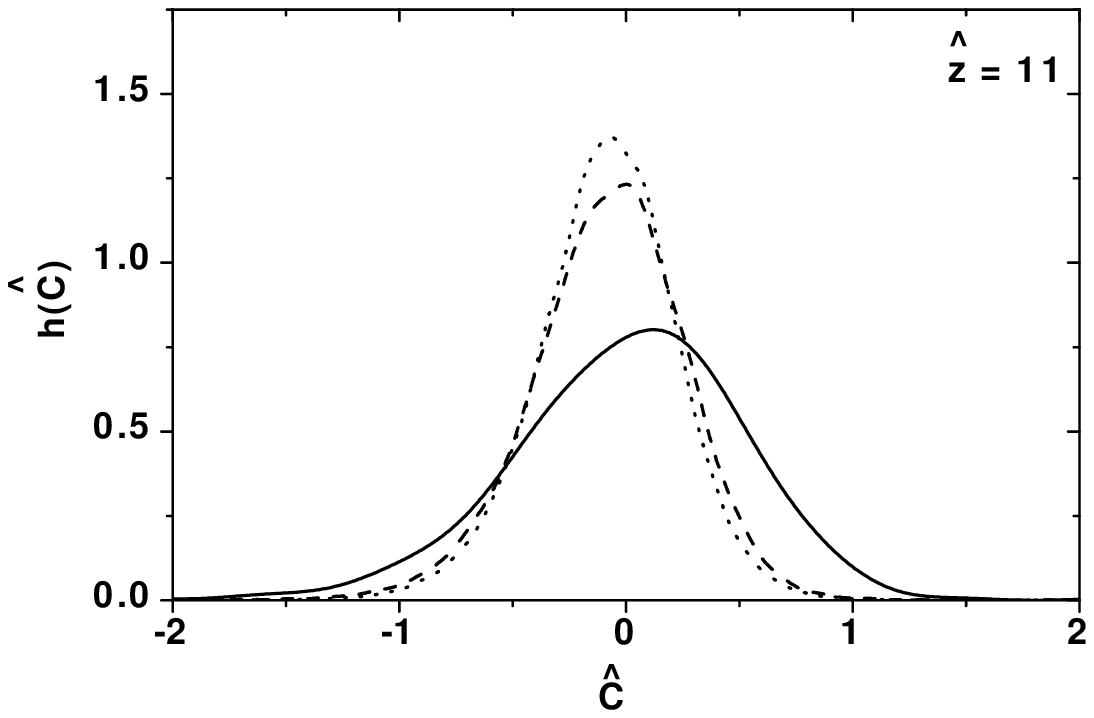}

\includegraphics[width=0.5\textwidth]{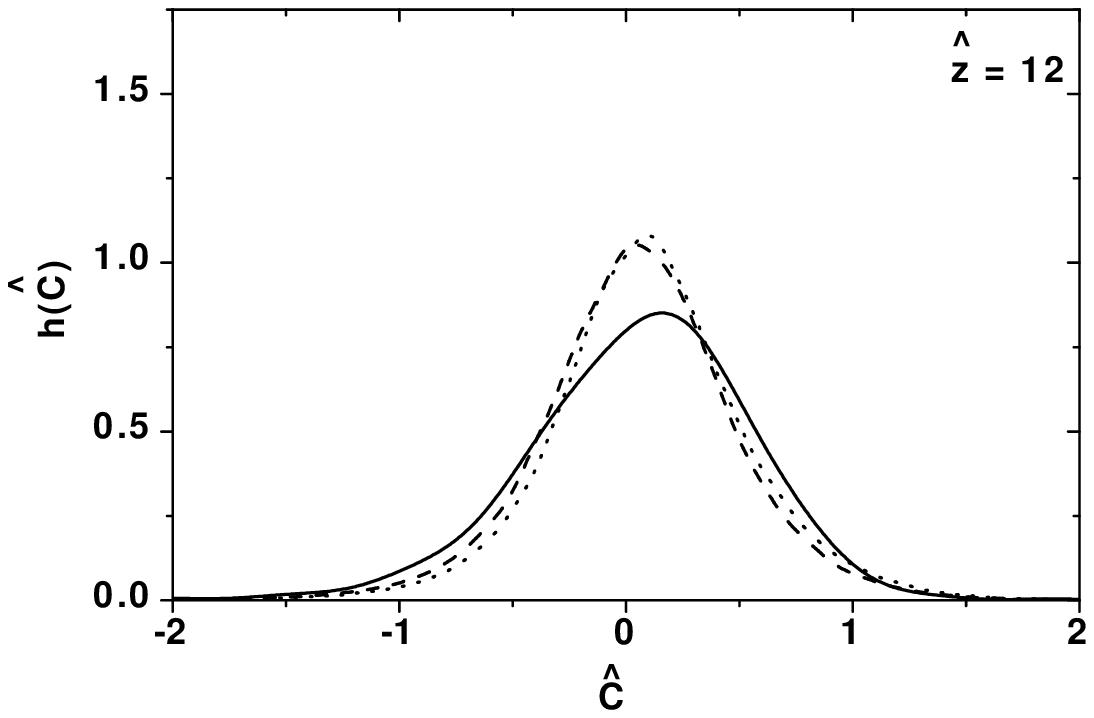}
\includegraphics[width=0.5\textwidth]{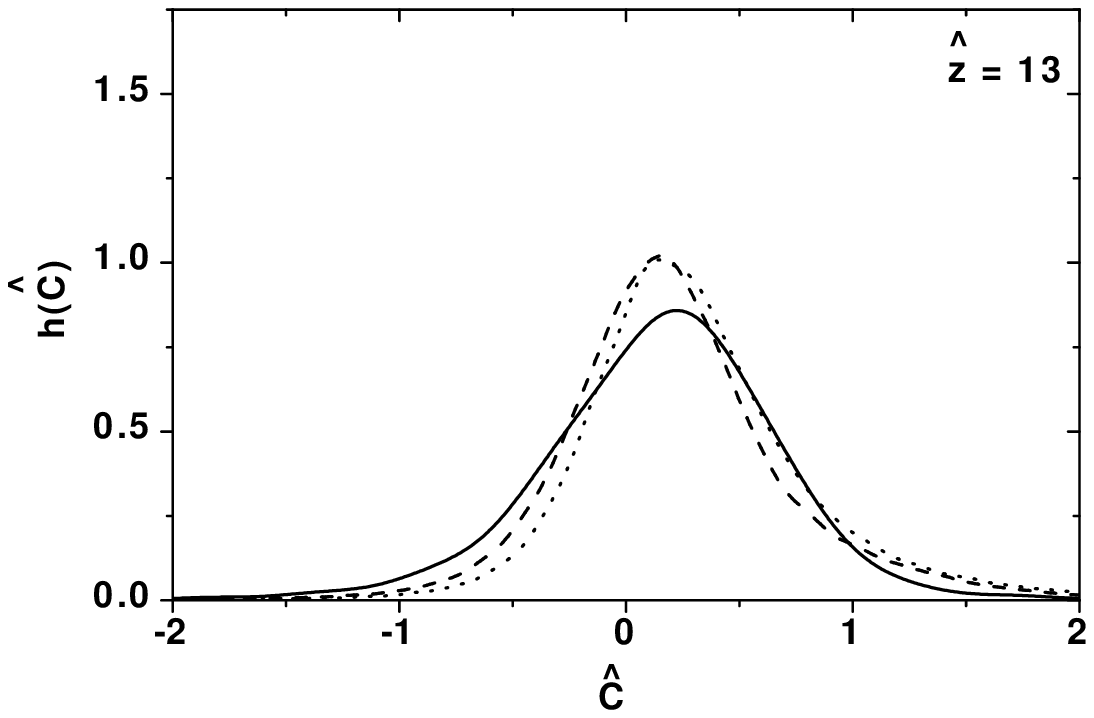}

\caption{
Normalized spectrum at different length of the undulator: $\hat{z}
=10-13$. Solid, dashed, and dotted lines correspond to the fundamental,
3rd and 5th harmonic, respectively
}
\label{fig:spectr135}

\end{figure}

When comparing radiation spectra, it is convenient to use the
normalized spectral density, $h(\hat{C})$, defined as

\begin{displaymath}
\int \limits_{-\infty}^{\infty}
\D \hat{C} h(\hat{C}) = 1 \ .
\end{displaymath}

\noindent The frequency deviation, $\Delta \omega$, from the nominal
value of $\omega_h$ can be recalculated as $\Delta \omega = - 2 \rho
\omega_h \hat{C}$. Since we consider the model of a long rectangular
bunch, the function $h(\hat{C})$ can be treated as the normalized
spectral density of both the radiation energy and the power.

Normalized envelope of the radiation spectrum and the first order time
correlation function are connected by the relation \cite{goodman}:

\begin{equation}
G(\Delta \omega ) =
\frac{1}{2\pi}
\int \limits^{\infty}_{-\infty}
d\tau g_1(\tau)\exp(-i\Delta\omega\tau) \ .
\label{spectr-env}
\end{equation}

The temporal structures of the radiation pulses (see
Fig.~\ref{fig:ptemp13}) are used for calculating the first order time
correlation function (see Fig~\ref{fig:acor135}). Then the radiation
spectra are reconstructed by Fourier transformation of the first order
time correlation function. Figure~\ref{fig:spectr135} shows evolution
of the radiation spectra of the SASE FEL radiation from the end of the
linear regime to saturation. Note that spectrum width of the higher
harmonics from SASE FEL differs significantly from that of incoherent
radiation. For the case of incoherent radiation relative spectrum
width, $\Delta \omega/\omega _h$ scales inversely proportional
to the harmonic number $h$ \cite{wiedemann}. One can see that situation
changes dramatically for the case when nonlenear harmonic genaration
process starts to be dominant. At saturation we find that relative
spectrum bandwidth becomes to be nearly the same for all odd harmonics.

\section{Summary}

In this paper we performed detailed study of the properties of the odd
harmonic of the radiation from SASE FEL. Universal formulae for
contribution of the higher odd harmonics to the FEL power for SASE FEL
operating at saturation are obtained. In the case of cold electron beam
these contributions are functions of the undulator parameter $K$ only.
General statistical properties of the odd harmonics of the SASE FEL
operating in saturation are as follows. Power of higher harmonics is
subjected to larger fluctuations than that of the fundamental one.
Probability distributions of the instantaneous power of higher
harmonics is close to the negative exponential distribution. The
coherence time at saturation falls inversely proportional to harmonic
number, and relative spectrum bandwidth remains constant with harmonic
number.


\clearpage

\begin{thebibliography}{99}

\bibitem{hg-1}
M. Schmitt and C. Elliot, Phys. Rev. A, 34(1986)6.

\bibitem{hg-2}
R.~Bonifacio, L.~De~Salvo, and P.~Pierini,
Nucl. Instr. Meth. A293(1990)627.

\bibitem{hg-2a}
W.M.~Fawley,
Proc. IEEE Part. Acc. Conf., 1995, p.219.

\bibitem{hg-3}
H. Freund, S. Biedron and S. Milton,
Nucl. Instr. Meth. A 445(2000)53.

\bibitem{hg-4}
H. Freund, S. Biedron and S. Milton,
IEEE J. Quant. Electr. 36(2000)275.

\bibitem{hg-5}
S. Biedron et al.,
Nucl. Instr. Meth. A 483(2002)94.

\bibitem{hg-6}
S. Biedron et al.,
Phys. Rev. ST 5(2002)030701.

\bibitem{kim-1}
Z. Huang and K. Kim, Phys. Rev. E, 62(2000)7295.

\bibitem{kim-2}
Z. Huang and K. Kim,
Nucl. Instr. Meth. A 475(2001)112.

\bibitem{hg-exp-1}
A. Tremaine et al., Phy. Rev. Lett. 88, 204801 (2002)

\bibitem{3rdharm-ttf}
W.~Brefeld et al.,
Nucl. Instr. Meth. A 507(2003)431.

\bibitem{fast}
E.L.~Saldin, E.A.~Schneidmiller, and M.V.~Yurkov,
Nucl. Instr. Meth. A 429(1999)233.

\bibitem{bon-rho}
R.~Bonifacio, C.~Pellegrini and L.M.~Narducci,
Opt. Commun. {\bf 50}(1984)373.

\bibitem{book}
E.L.~Saldin, E.A.~Schneidmiller, and M.V.~Yurkov, The Physics of Free
Electron Lasers, Springer-Verlag, Berlin, 1999.

\bibitem{fawley-loading}
W.M.~Fawley,
Phys. Rev. STAB, 5(2002)070701.

\bibitem{statistics-oc}
E.L.~Saldin, E.A.~Schneidmiller, and M.V.~Yurkov,
Opt. Commun. 148(1998)383.

\bibitem{attofel}
E.L.~Saldin, E.A.~Schneidmiller, and M.V.~Yurkov,
Opt. Commun. 212(2002)377.

\bibitem{goodman}
J. Goodman, Statistical Optics, Willey, New York, 1985.

\bibitem{wiedemann}
H. Wiedemann, Synchrotron Radiation,
Springer-Verlag, Berlin, 2003.

\end{thebibliography}
\end{document}